\documentclass[aip,amsmath,amssymb,reprint]{revtex4-2}
\usepackage{graphicx}
\usepackage{dcolumn}
\usepackage{bm}
\usepackage[utf8]{inputenc}
\usepackage[T1]{fontenc}
\usepackage{mathptmx}
\usepackage{etoolbox}
\usepackage{amsmath}
\usepackage{caption}
\usepackage{siunitx}
\usepackage{subcaption}
\usepackage{hyperref}
\usepackage{float}
\graphicspath{{./images/}}

\makeatletter
\def\@email#1#2{
 \endgroup
 \patchcmd{\titleblock@produce}
  {\frontmatter@RRAPformat}
  {\frontmatter@RRAPformat{\produce@RRAP{*#1\href{mailto:#2}{#2}}}\frontmatter@RRAPformat}
  {}{}
}
\makeatother
\draft
\begin{document}
\nocite{*}
\preprint{AIP/123-QED}

\title[A. Karmali \& J-F. Milette - PHY3904 Applied Physics Laboratory II]{Droplet Interferometry: A Schlick Way to Consider Interfacial Energetics}
\author{A. Karmali}
 \email{jmile020@uottawa.ca}
 \affiliation{University of Ottawa, Department of Physics}
 
\author{J-F. Milette}
 \email{akarm060@uottawa.ca}

\date{\today}

\begin{abstract}

\textbf{Abstract:} We verify the use of an evaporating sessile water droplet as a source of dynamic interference fringes in a Fizeau-like interferometer. Experimentally-obtained interference patterns are compared with those produced by a geometrical optics-based computational model to demonstrate the potential for classical optical theory to enhance the analysis of interfacial energetics. A detailed description of the process taken to optimize fringe visibility is presented, and a comparison is made between various droplet substrates. Silicon-based substrates appear to be superior than glass-based substrates in their ability to image a clear dynamic interference pattern.

\end{abstract}

\maketitle

\section{\label{sec:level1}Introduction}
\subsection{Optical Theory}
\noindent
It has been well-established that in order for the spatial overlap of electromagnetic waves to produce a sustained optical interference pattern, the combining waves must originate from a coherent, monochromatic light source. In other words, interfering wavefronts must be characterized by a single wavelength (\textit{$\lambda$}), and must have a well-defined phase evolution as they travel through the optical system. These are the conditions under which the superposition principle is valid, a mathematical tool that renders the calculation of the resulting wave amplitude at a particular point in space to be the algebraic sum of the component wave amplitudes in that region. Since the intensity of a light wave is proportional to its amplitude squared, the intensity resulting from the superposition of two monochromatic light waves can be obtained by squaring the superposition of the component wave amplitudes. With the help of trigonometry, the following equation may be obtained:

\begin{equation}\label{eq:resulting intensity}
    \boxed{I_{Total} = I_1 + I_2 +2\sqrt{I_1 I_2}\;\cos\;(\frac{2 \pi}{\lambda}\delta)}
\end{equation}
\\
\noindent
While the superposition principle implies a continuum of possible resulting intensities, a focus is often placed on two limiting cases. The case of destructive interference corresponds to the minimum resulting intensity, whereas constructive interference is attributed to the maximum resulting intensity. Optical interference patterns are characterized by alternating regions of these limiting cases, and the specific patterning of these bright and dark regions is largely determined by the optical path length differences ($\delta$) induced by the optical system at hand. When these alternating regions are viewed on an observation screen, the central fringe is denoted as the 0th order. A nonzero integer order (\textit{m}) is then attributed to each similar fringe on the observation screen, moving outwards from the 0th order fringe. The optical path length difference required to create these limiting cases is given by equations \ref{eq:cond_destr} and \ref{eq:cond_const}, which show that higher order fringes represent greater optical path length differences between component waves.

\begin{equation}\label{eq:cond_destr}
    \boxed{\delta=(m+1/2)\lambda}  \hspace{0.5 in}  \text{(Destructive)}
\end{equation}

\begin{equation}\label{eq:cond_const}
    \boxed{\delta=m\lambda} \hspace{0.5 in} \text{(Constructive)}
\end{equation}

\begin{figure}
    \centering
    \includegraphics[scale=1.8]{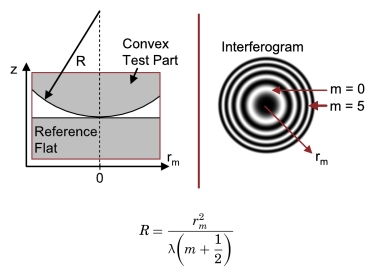}
    \caption{A so-called Fizeau interferometer composed of a plano-convex lens having a radius of curvature \textit{R}, with a reflective plate below. Fizeau fringes are also depicted with a model relating the lens' radius of curvature to the fringe order \textit{m}, wavelength \textit{$\lambda$}, and radial distance from the central dark spot \textit{r\textsubscript{m}}. Adapted from \cite{Fizeau}.}
    \label{fig:fizeau}
\end{figure}

\noindent
Interferometers are a class of instruments that produce optical interference patterns by subjecting beams of coherent light to varying optical path lengths in order to relate microscopic changes in path length to macroscopic changes in the imaged interference pattern. Since the wavelength of visible light is on the order of a couple hundred nanometers, interferometry based on monochromatic visible light sources like He-Ne lasers (632.8 nm) is well-suited for detecting nanoscale displacements of the components in the optical system.  
\\
\\
\noindent
 The Michelson interferometer typically produces an interference pattern by splitting coherent light with a beam splitter into two beams that travel different optical path lengths before being recombined. The discrepancy in optical path length ($\delta$) induces a relative phase difference between the beams, and the resulting interference pattern is decorated by bright and dark fringes whose placement on the observation screen obeys equations \ref{eq:cond_destr} and \ref{eq:cond_const}.  If the difference in optical path length travelled prior to recombination is held fixed, a static interference pattern is generated on the observation screen, but if the path length difference is made to vary, say by translating a reflective surface in the optical system, the resulting change in relative phase between recombining beams gives rise to a dynamic fringe pattern whose motion can be related to the underlying change in optical path length.
\\
\\
\noindent
Thin film interference occurs when a monochromatic light wave is incident on a film whose thickness is on the order of the light's wavelength.  Light rays reflecting off of the various interfaces of the thin film may recombine at some image point, producing an interference pattern. Since reflection off of a medium with a higher index of refraction results in a change in path length difference of $ \lambda/2$, not only the thickness of the film but also the indices of refraction of the incident medium, the thin film and the underlying substrate determine the relative path length difference between light rays recombining at the image point. Depending on the optical boundaries involved, many reflections may occur within the thin film, and an interference pattern can arise similar to that seen in a Fabry-Perot interferometer. 
\\ 
\\
\noindent
Haidinger fringes, or fringes of equal inclination, describe an interference pattern for which each region of constructive or destructive interference is associated to light with a particular angle of incidence. As the angle of incidence changes, the transmitted light takes a different path through the thin film, and the difference between this optical path length and that of the purely reflected beam gives rise to Haidinger fringes. Fizeau fringes, or fringes of equal thickness, are attributed to interference patterns which are generated by reflections off a thin film with varying thickness. The varied thickness creates fringes in an analogous way to Haidinger fringes. 
\\
\\
\noindent
Newton's rings are a type of interference pattern composed of concentric rings of bright and dark regions. A common way to obtain this pattern is through the placement of a plano-convex lens on a planar glass surface, provided that the lens' radius of curvature is much larger than the radial distances along the lens under consideration. Geometrically, this system is relevant to the case of Fizeau fringes, as the curvature of the lens allows a dielectric layer of varying thickness between the lens and the glass plate (see FIG \ref{fig:fizeau}). Some incident light rays reflect off of the curved glass surface, while others pass through the dielectric layer, and the recombination of the rays produces an interference pattern on an observation screen placed above the plano-convex lens. The circular symmetry of this fringe pattern can also be used to assess the surface regularity of a polished lens; an optical element with a symmetric surface curvature will produce Newton's rings, and the distortion of these concentric rings can indicate an asymmetry introduced during the polishing process. As seen in FIG \ref{fig:fizeau}, the central spot in Newton's rings is typically a region of destructive interference, a result of the $\lambda/2$-shift experienced during reflection at the second dielectric-glass boundary relative to immediate reflection off of the curved glass interface. The convention for the fringe order used in FIG \ref{fig:fizeau}'s model is to assign an integer value to each fringe, bright or dark, starting with the first bright fringe. 
\\

\noindent
Far away from their source, electromagnetic waves are well approximated by plane waves: equally-spaced propagating surfaces of constant phase. Geometrical optics interprets this approximation as rays normal to the propagating wavefronts, greatly simplifying the description of light-matter interactions with objects much larger than the wavelength of light. Thus, the boundary conditions of electromagnetic waves at an interface can be turned into relatively simple geometrical arguments. The plane of incidence defines a plane which contains the incident ray, the reflected ray, the normal to the interface, as well as the transmitted ray. For the case of specular reflection, a light ray with a particular angle of incidence ($\theta_i$) measured with respect to the normal has a reflected component emerging with an angle of reflection ($\theta_R$) such that $\theta_I=\theta_R$. If this incident ray is initially travelling in a medium with refractive index $n_i$, Snell's law states that the portion transmitted to the next medium $n_T$ propagates at an angle $\theta_T$ according to the following:

\begin{equation}\label{eq:snell}
    \boxed{n_i \sin(\theta_i)=n_T \sin(\theta_T)}
\end{equation}

\noindent
To obey energy conservation, whatever energy is transported by the incident ray will have to be accounted for in the energy carried by the reflected and transmitted rays. The energy contained in an electromagnetic wave can be expressed in terms of the amplitude of its electric field component, and the application of boundary conditions at the interface gives rise to Fresnel's equations. These equations relate the amplitude of the electric field of the incident beam to that of either the reflected or transmitted beams, and this is done separately for both linear polarization states of the electric field vector. For the case of p-polarized light, where the electric field is parallel to the plane of incidence, the amplitude of the incident field ($E_{p,i}$) is related to the reflected ($E_{p,R}$) and transmitted fields ($E_{p,T}$) by the following:

\begin{equation}\label{eq:P-reflection}
    \boxed{r_{\parallel}=\frac{E_{p,R}}{E_{p,i}}=\cfrac{n_i\cos(\theta_T)-n_T\cos(\theta_i)}{n_i\cos(\theta_T)+n_T\cos(\theta_i)}}
\end{equation}

\begin{equation}\label{eq:P-transmission}
    \boxed{t_{\parallel}=\frac{E_{p,T}}{E_{p,i}}=\cfrac{2n_i\cos(\theta_i)}{n_i\cos(\theta_T)+n_T\cos(\theta_i)}}
\end{equation}

\noindent
And for s-polarized light, where the electric field is perpendicular to the plane of incidence:

\begin{equation}\label{eq:S-reflection}
    \boxed{r_{\perp}=\frac{E_{s,R}}{E_{s,i}}=\cfrac{n_i\cos(\theta_i)-n_T\cos(\theta_T)}{n_i\cos(\theta_I)+n_T\cos(\theta_T)}}
\end{equation}

\begin{equation}\label{eq:S-transmission}
    \boxed{t_{\perp}=\frac{E_{s,T}}{E_{s,i}}=\cfrac{2n_i\cos(\theta_i)}{n_i\cos(\theta_i)+n_T\cos(\theta_T)}}
\end{equation}
\\

\noindent
 This version of Fresnel's equations assumes non-magnetic media with permeability $\mu=\mu_o$ and permittivity $\epsilon$. The index of refraction characterizes the speed of light in a medium and it is given by equation \ref{index}. 

 \begin{equation}\label{index}
     \boxed{n=\sqrt{\cfrac{\epsilon}{\epsilon_o}}}
 \end{equation}

\noindent
The refractive index also serves as a conversion factor between the physical path length in an optical system (\textit{d}) and the corresponding optical path length (\textit{L}), as shown in equation \ref{OPL}. This conversion allows for the quantification of the optical path length difference between two light waves ($\delta$) required for the mathematical treatment of optical interference. 

\begin{equation}\label{OPL}
    \boxed{L=nd
    ,\quad \delta=L_1 - L_2}
\end{equation}
\\
\noindent
 As mentioned, the power delivered per unit area (\textit{I}) is proportional to the square of the electric field amplitude ($E_o$), with the proportionality factor given by the speed of light in free space (\textit{c}), the refractive index (\textit{n}), and the permittivity of the optical medium ($\epsilon$). This relationship is shown by equation \ref{eq:intensity}, which can be conveniently rearranged using equation \ref{index}.

 \begin{equation}\label{eq:intensity}
    \boxed{I=\frac{c}{2n}\epsilon E_o^2=\frac{cn\epsilon_o}{2} E_o^2}
 \end{equation}
\\
\noindent
 Although the electric field amplitudes of each beam may not be known, Fresnel's equations provides the ratios between them, and so the ratio between intensities can also be found explicitly. The reflectance (\textit{R}) and transmittance (\textit{T}) denote the proportion of incident beam energy carried by the reflected and transmitted beams, respectively. However, it is still the case that each linear polarization state must be handled independently.

\begin{equation}\label{eq:reflectance}
    \boxed{R_p=\cfrac{I_{p,R}}{I_{p,i}}=(r_{\parallel})^2
    ,\quad R_s=\cfrac{I_{s,R}}{I_{s,i}}=(r_{\perp})^2}
\end{equation}

\begin{equation}\label{eq:transmittance}
    \boxed{T_p=\cfrac{I_{p,T}}{I_{p,i}}=\cfrac{n_T}{n_i}\;(t_{\parallel})^2
    ,\quad T_s=\cfrac{I_{s,T}}{I_{s,i}}=\cfrac{n_T}{n_i}\;(t_{\perp})^2}
\end{equation}

\begin{equation}\label{eq:energyconservation}
    \boxed{R_p+T_p=1
    ,\quad R_s+T_s=1}
\end{equation}
\\
\noindent
 When a light source is said to be unpolarized, there is no preferred orientation for the electric field vector, and so it is randomly oriented in a direction perpendicular to the direction of propagation. The effective reflectance (\textit{R}) and transmittance (\textit{T}) for unpolarized light can be obtained by averaging the contributions from both linear polarization states.

\begin{equation}\label{eq:effreflectance}
    \boxed{R = \frac{1}{2}(R_p+R_s)}
\end{equation}

\begin{equation}\label{eq:efftransmittance}
    \boxed{T = \frac{1}{2}(T_p+T_s)}
\end{equation}
\\
\noindent
Note that taking the average like this is only appropriate for isotropic materials, which have the same optical properties in all directions. For anisotropic materials, which do not possess this symmetry, the effective reflectance and transmittance will depend on the orientation of the material with respect to the instantaneous polarization state of the incident light.
\\
\\
Schlick's approximation is a computationally-efficient model for evaluating the reflectance in the case of specular reflection off of a non-conducting interface\cite{schlick}. If the incident angle and refractive indices are known, Schlick's approximation can be utilized:

\begin{equation}\label{eq:schlick}
    \boxed{R = R_0 + (1-R_0)(1-\cos{\theta_i})^5 ,\quad R_0 = (\frac{n_i-n_T}{n_i+n_T})^2 }
\end{equation}

\subsection{Mathematical Theory}

%Je crois qu'il faudrait preciser le choix de direction utiliser dans la formule pour N(t) 

\noindent
 A parametric curve can be described symbolically as: $\pmb{\vec{r}}(t) = \pmb{\vec{r}}(0) + \pmb{\vec{a}}(t)$, where $\pmb{\vec{r}}(t)$ is a point on the curve evaluated at parameter $t$, $\pmb{\vec{a}}(t)$ gives the "direction" of the curve at this point, and $\pmb{\vec{r}}(0)$ is a vector to the starting point of the curve. For a ray in 2 dimensions, $\pmb{\vec{a}}(t)$ is a linear function of the form $\pmb{\vec{a}}(t) = [a_0 t,\; a_1 t]$, for real numbers $a_0$ and $a_1$. For an ellipse centered at the origin, the parametric equation will be $\pmb{\vec{r}}(t) = [a\cos(t),\; b\sin(t)]$, where $a$ and $b$ are real numbers, and $0 < t \leq 2 \pi$. Here, $a$ defines the radius of the horizontal axis of the ellipse, while $b$ defines the radius of the vertical axis which can also be written as $R_x$ and $R_y$, respectively.
\\
\\
This parametric form is useful for calculating the angle between two rays, as the angle between two vectors can be found using the dot product, shown in equation \ref{dotproduct}. In the case of a line, the vectors $\pmb{\vec{A}}$ and $\pmb{\vec{B}}$ is given by the $\pmb{\vec{a}}(t)$ of the line's parametric equation, since only the direction of the lines are necessary to find the angle $\theta$.

\begin{equation}\label{dotproduct}
\boxed{\pmb{\vec{A}} \cdot \pmb{\vec{B}} = \lVert \pmb{\vec{A}} \rVert \; \lVert \pmb{\vec{B}} \rVert \; \cos(\theta)}
\end{equation}
\\

%Since the Snell's law (equation \ref{eq:snell}) %requires the angle between the normal and the %incident ray, we need to find the normal at %each point on the ellipse. To do this, we will %use calculus. 

\noindent
The tangent at a particular point of the parametric curve, $\pmb{\vec{T}}(t)$, can be found by taking the derivative and normalizing it to get a unit vector. Finding the normal in the 2D case involves swapping the two coordinates of the tangent vector, as well as switching the sign of one coordinate, depending on whether the normal direction $\pmb{\vec{N}}(t)$ points inwards or outwards from the center of the ellipse. An outward-facing normal can be represented as follows:

\begin{equation}
\begin{aligned}
    \pmb{\vec{T}}(t) =\frac{\pmb{\vec{r}}'(t)}{\lVert \pmb{\vec{r}}'(t)  \rVert} = \frac{1}{\lVert \pmb{\vec{r}}'(t)  \rVert}[r_x'(t),r_y'(t)]  \\ \pmb{\vec{N}}(t) = \frac{1}{\lVert \pmb{\vec{r}}'(t)  \rVert}[r_y'(t),-r_x'(t)]
    \end{aligned}
\end{equation}

\noindent
\textbf{Objective:} The aim of the following section is to outline the procedure undertaken to image a dynamic fringe pattern through the evaporation of a sessile water droplet. Many experimental adjustments are presented, partly elucidating various characteristics of this interferometry-based optical system. A computational approach using the optical and mathematical theory outlined above is then performed to compare the results of simulation with those produced experimentally.  

\section{Methodology}

\subsection{General Approach}
\begin{figure}[h] 
    \centerline{\includegraphics[angle=270,scale=0.4]{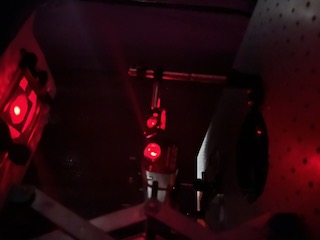}}
    \caption{Example of experimental setup: a Metrologic ML868 He-Ne laser is incident on a 50/50 beam splitter, which is positioned to direct laser light onto a water droplet. Here, the reflective surface beneath the droplet is a silicon wafer, and light is sent back up through the beam splitter to the observation screen placed above.}
    \label{fig:setup}
\end{figure}

\noindent
Motivated by the experimental setup used by G-L. Ngo et al\cite{template}, a water droplet is placed on a reflective surface (substrate), and a laser is made to shine directly onto the droplet, with the incident beam oriented perpendicularly to the reflective surface. The reflected light is then sent towards an observation screen for viewing, as shown in FIG \ref{fig:setup}. The trajectory of the laser light throughout this optical system is also simulated in Python using ray optics, so that the computed interference pattern may be compared with that obtained at the physical observation screen.

\subsection{Simulated Model}

\begin{figure}[H] 
    \centerline{\includegraphics[scale=0.21]{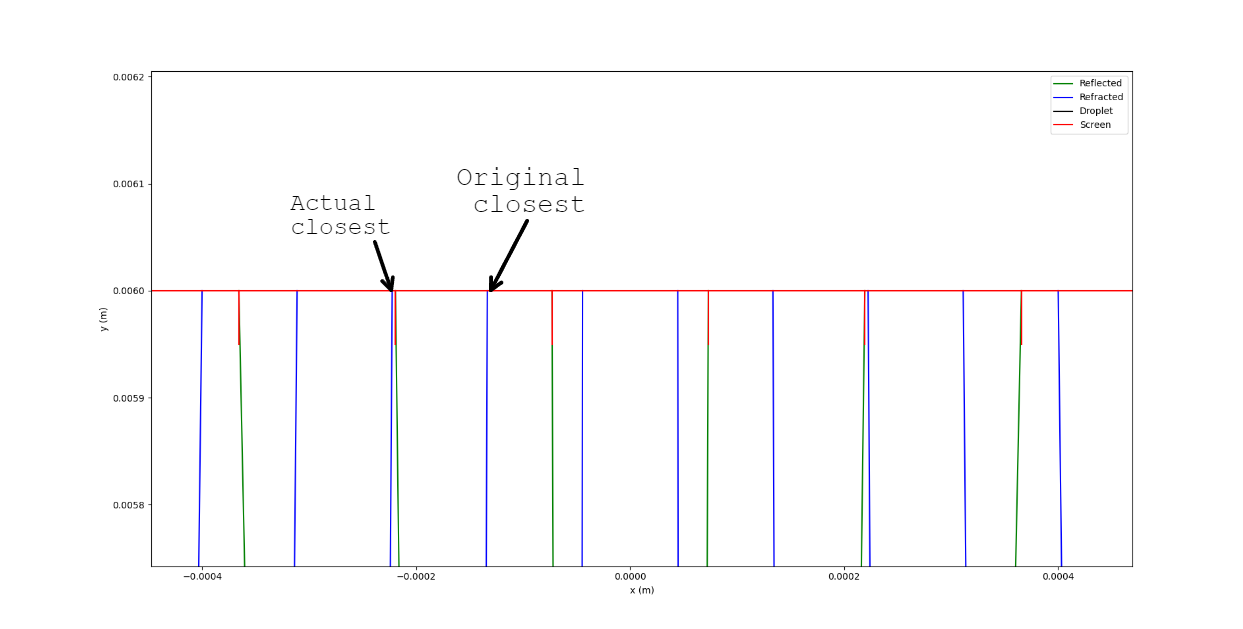}}
    \caption{When there is no shift, we can be in a situation where the "closest" refracted (blue) ray inside the associated pixel of the reflected (green) ray is not actually the closest.}
    \label{fig:annotate}
\end{figure}

\begin{figure}[H] 
    \centerline{\includegraphics[scale=0.21]{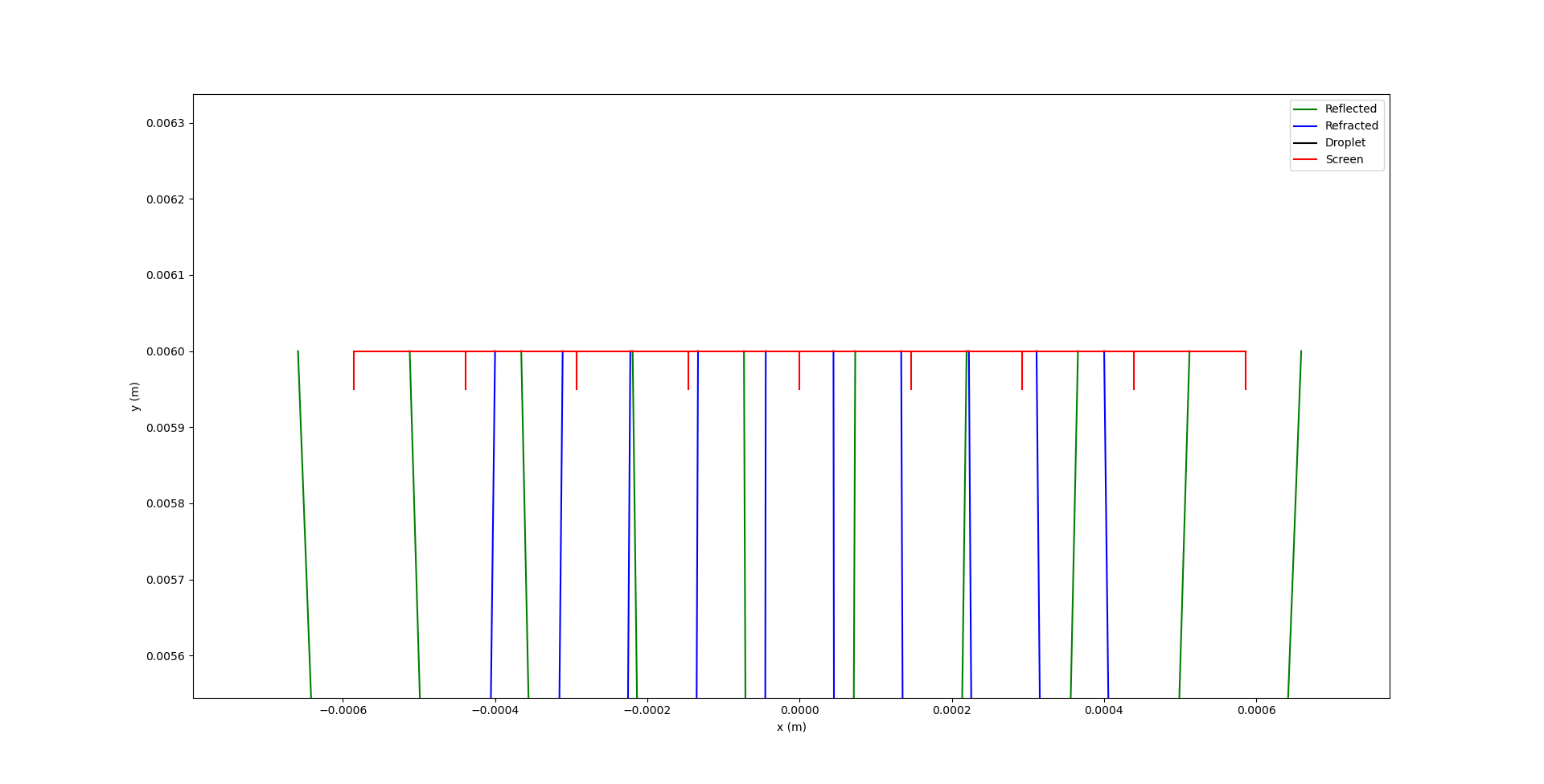}}
    \caption{Shifting each pixel (red vertical lines) by half its width ensures that the closest refracted ray always falls inside the reflected ray's pixel. This may not always be true for the outermost pixels, but these are not the pixels involved in detecting the desired interference pattern.}
    \label{fig:shift}
\end{figure}

\noindent
The incident laser beam can be broken up into a large number of rays that each obey the predictions of geometrical optics. Since the ML868 laser outputs a Gaussian beam with a diameter of 0.66 mm, not all rays will carry the same light intensity. Rays from the center of the beam will possess maximum intensity, and this intensity will decay to $1/e^2$ at a radial distance of 0.33 mm. Each incident ray can be broken down into four components upon interaction with the droplet's surface: one  component being immediately reflected off of the air-droplet boundary and three components to describe refraction through the droplet before reaching the screen. For the latter case, one component describes the entry into the droplet, the second gives the resulting reflection off of the substrate, and the third is the portion of the ray that emerges from the droplet and travels to the observation screen. Each of these components are computed using the law of specular reflection and the Snell's law (equation \ref{eq:snell}) at each optical boundary. When light reflects off of the air-droplet interface, a path length difference of $\lambda/2$ is acquired. Similarly, whether the substrate is glass or silicon, the refracted ray also experiences the same shift after reflecting off of the substrate, so these phase shifts can be ignored in the simulation. 
\\
\\
While each ray possesses an initial intensity dictated by its radial distance from the beam center, the intensity contributed to the observation screen is computed using three distinct models which consider loss of light at each optical boundary. The first model simply splits the incident intensity equally between the resulting reflected and transmitted beam. The second model makes use of Schlick's approximation (equation \ref{eq:schlick}) for the reflectance and transmittance ($T=1-R$) at each boundary. Lastly, equations \ref{eq:reflectance} - \ref{eq:energyconservation} are used to build the Fresnel model for light loss. The water droplet is assumed to be a non-magnetic, non-conducting homogeneous optical medium.
\\
\\
\noindent
However, since interference is a wave-like property of light and is not naturally described by ray optics, we must devise a way to calculate the interference of rays at the observation screen. Our choice was to compute the intensity contribution of both a reflected ray and a refracted ray combining at the observation screen using equation \ref{eq:resulting intensity}. Note that these two interfering rays generally won't originate from the same incident ray like in the case of the Michelson interferometer. The goal is to associate each pair of interfering reflected/refracted rays with one "pixel" on the screen, but we must make sure that the interference calculation considers only a pair of rays at each pixel. If we assign a pixel to each reflected ray, then depending on the height of the observation screen relative to the focal length of the droplet, more than one refracted ray may converge into each pixel (see FIG \ref{fig:annotate}). To remediate this, only the interference between a reflected ray and the closest refracted ray will be considered. The width of a pixel will then vary based on the distance between its associated reflected ray and the closest reflected ray. By shifting the boundary of each pixel by half its width, the reflected rays are in the middle of their pixel, facilitating the identification of the closest refracted ray (see FIG \ref{fig:shift}). Once we identify each pair of rays, we take into account their trajectories and make use of equation \ref{OPL} to compute the optical path length differences. The intensity detected at the pixel may then be computed using equation \ref{eq:resulting intensity}.

\noindent
To simulate the evaporation of the droplet over time, two different models were used. The first model involves reducing both $R_x$ (a) and $R_y$ (b) at the same rate, amounting to a reduction in droplet size while maintaining the same elliptical structure. The second model involves reducing $R_y$, while keeping $R_x$ fixed, making for a fixed droplet width that only loses height over time.

\begin{figure}[H] 
    \centerline{\includegraphics[scale=0.19]{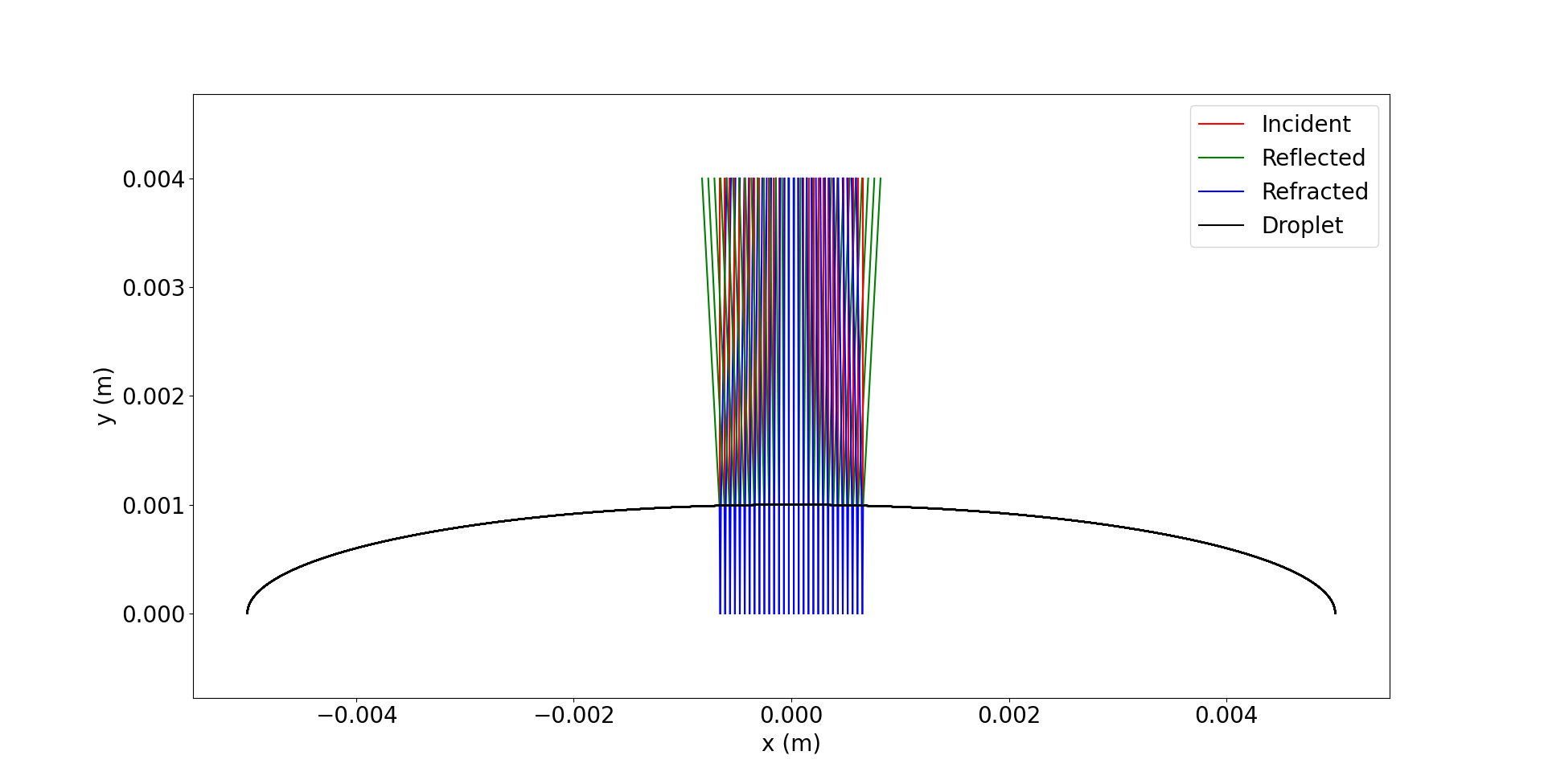}}
    \caption{Our model of the incident laser light striking the water droplet surface. The gaussian beam is split into 30 rays here for better viewing.}
    \label{fig:Setup_Fresnel_testing}
\end{figure}

\subsection{Obtaining and Optimizating the Desired Fringe Pattern}

\begin{figure}[h]
\centering
\includegraphics[height=1.5in, width=1.5in]{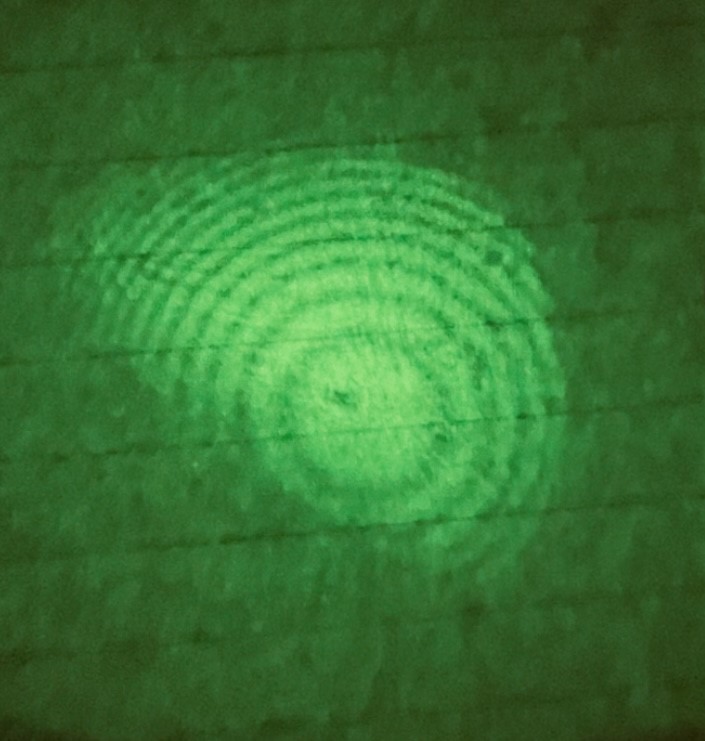}
\includegraphics[height=1.5in, width=1.5in]{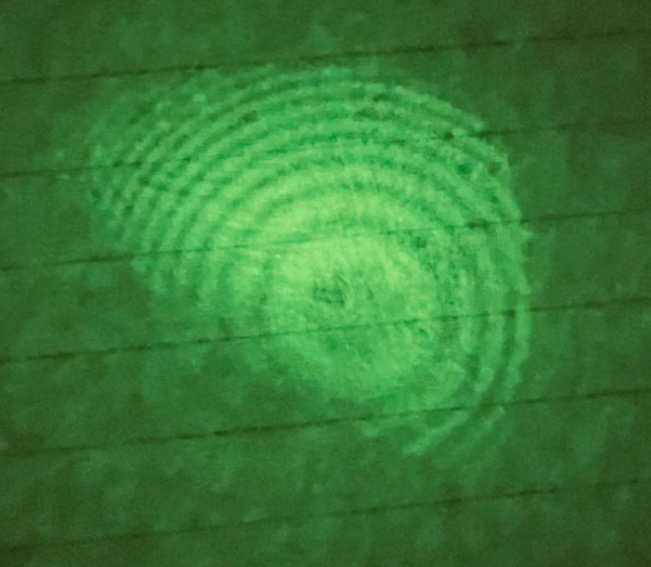}
\caption{Two screenshots taken from a video recording the slow dynamic interference pattern arising when the hot plate is added to the setup. These two pictures depict the dynamic fringe pattern as a transition from constructive interference in the center (left) to destructive interference (right). Part of the pattern is cut off due to poor alignment.}
\label{fig:thermalpattern}
\end{figure}

\noindent
As a coherent source of monochromatic light, an old He-Ne laser was successfully used to produce a dynamic fringe pattern on the observation screen, but its unstable light output made it difficult to clearly image these patterns, limiting our ability to probe their origin. The unstable laser output problem was fixed by switching to a 530 nm solid-state green laser, but the exact model is not known. To direct the laser light towards the droplet at normal incidence, a plate beam splitter was placed at an angle of 45$^{\circ}$ relative to the incident laser light; mirrors can also accomplish this, but in the setup depicted in FIG \ref{fig:setup}, replacing the beam splitter with a mirror would block the reflected light from reaching the observation screen above, and so an interference pattern would not be imaged. Beam splitters are usually polished in such a way as to reflect one portion of incident light while transmitting the other. This property is useful since it allows for the redirection of light towards the droplet while simultaneously allowing the light reflected back off of the droplet to continue unimpeded to the observation screen. Further, beam splitters can be engineered in order to customize this ratio of reflected to transmitted light; in this case, the ratio is simply 50/50. However, as previously mentioned in the context of the Fizeau interferometer in FIG \ref{fig:fizeau}, the polishing process of optical components like beam splitters can produce the phenomenon of Newton's rings. While this pattern may provide insights to asymmetries in the optical component's surface, it greatly reduces the visibility of the desired dynamic fringe pattern.  A beam splitter cube was tested in comparison, but a corner of the cube was damaged, making the laser alignment process more difficult. Thus, the plate beam splitter will be kept, despite the reduced visibility created by Newton's rings. 
 \\
 \\
\noindent
 A hot plate was acquired as a means to increase the evaporation rate of the water droplet. Although the pattern obtained with the first laser in standard ambient temperature and pressure conditions already had a fringe passing rate of about 2 fringes per second, it was not confirmed that the imaged pattern was the direct result of the water droplet evaporating. With the heater on, if a dynamic interference pattern with an even faster fringe passing rate than the first time is observed, it may then be hypothesized that the interference patterns imaged so far were indeed due to the evaporation of the water droplet. Using a squirt bottle, a water droplet was placed on a glass plate, which itself rested on a hot plate set to low heat to avoid cracking the glass. As a result, a dynamic interference pattern was produced on the observation screen (see FIG \ref{fig:thermalpattern}), but this time with a fringe passing rate of around 0.1 fringes per  second. Given that the fringe passing rate apparently slowed down when heat was applied, it can be said that these patterns are not arising from the same mechanism. The hot plate was abandoned because it was hard to standardize trials without a temperature gauge, and the application of heat risked damaging our substrate material.

\noindent
While the solid-state green laser solved the problem of unstable output intensity, it was soon needed elsewhere. Consequently, a Class II ML868 He-Ne laser manufactured by Metrologic was acquired and kept for the rest of the experiment. This laser has a radiant power output of 0.6-0.95 mW, and outputs a $TEM_{00}$ mode with a random linear polarization state\cite{Laser}. The specs sheet states that the beam diameter is about 0.66 mm with a peak wavelength of 632.8 nm. The static fringe pattern presumed to be a case of Newton's rings is still present with this new laser, supporting the notion that it is a result of the beam splitter's surface. 
\\
\\
\noindent
After some time with the He-Ne laser, it became relatively easy to obtain dynamic fringe patterns as a result of aligning the diffuse spot on the screen from reflection off the droplet with the spot caused by the refracted light. As long as the system is aligned such that the reflected light passes right back through the beam splitter, the desired pattern can be obtained simply by placing the laser spot near the center of the water droplet. However, since the squirt bottle did not produce a uniform droplet size, the alignment needed for proper viewing of the dynamic fringe pattern varied from trial to trial. The squirt bottle was replaced with an actual pipette so that the droplet size could be more easily standardized. 

\begin{figure}[h]
\centering
\includegraphics[scale=1.8]{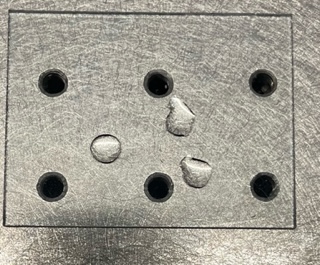}
\caption{Droplets formed with a pipette. The droplet on the left is an example of our standardized droplet volume (about 5 mm diameter), whereas the two droplets on the right have partially collapsed due to the microscopic conditions of the glass slide's surface.}
\label{fig:glassdroplet}
\end{figure}

\begin{figure}[h]
\centering
\includegraphics[height=1.5in, width=1.5in]{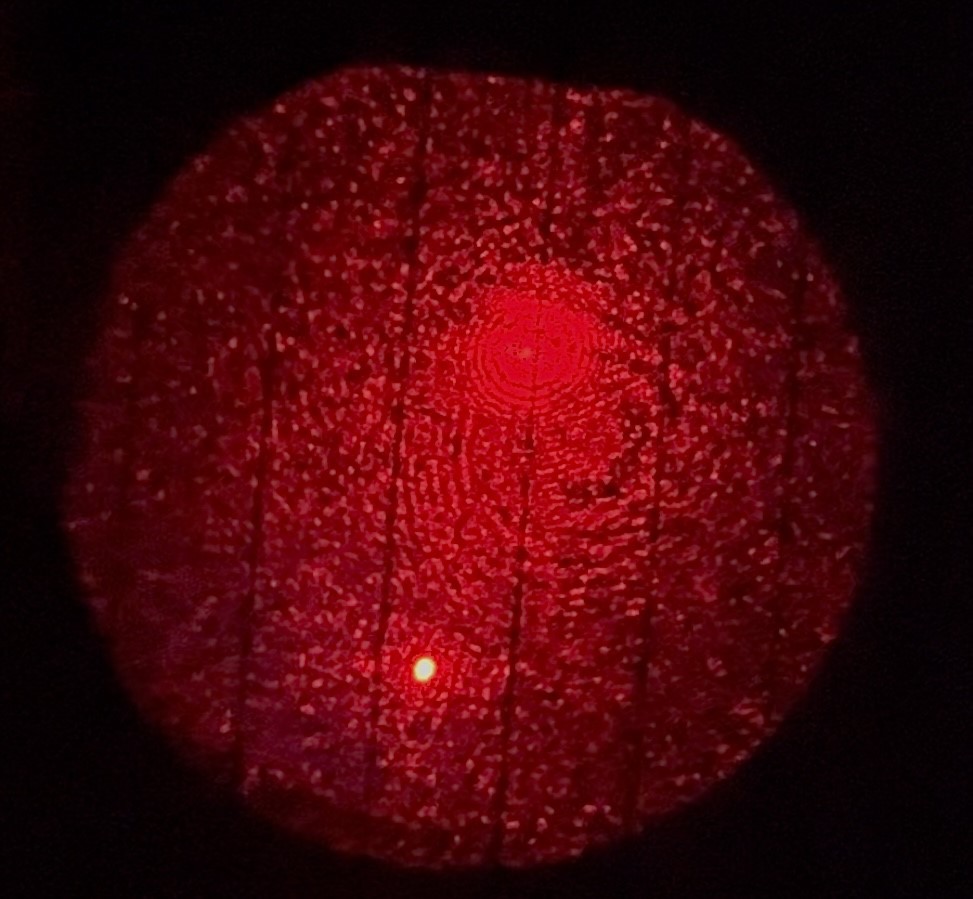}
\includegraphics[height=1.5in, width=1.5in]{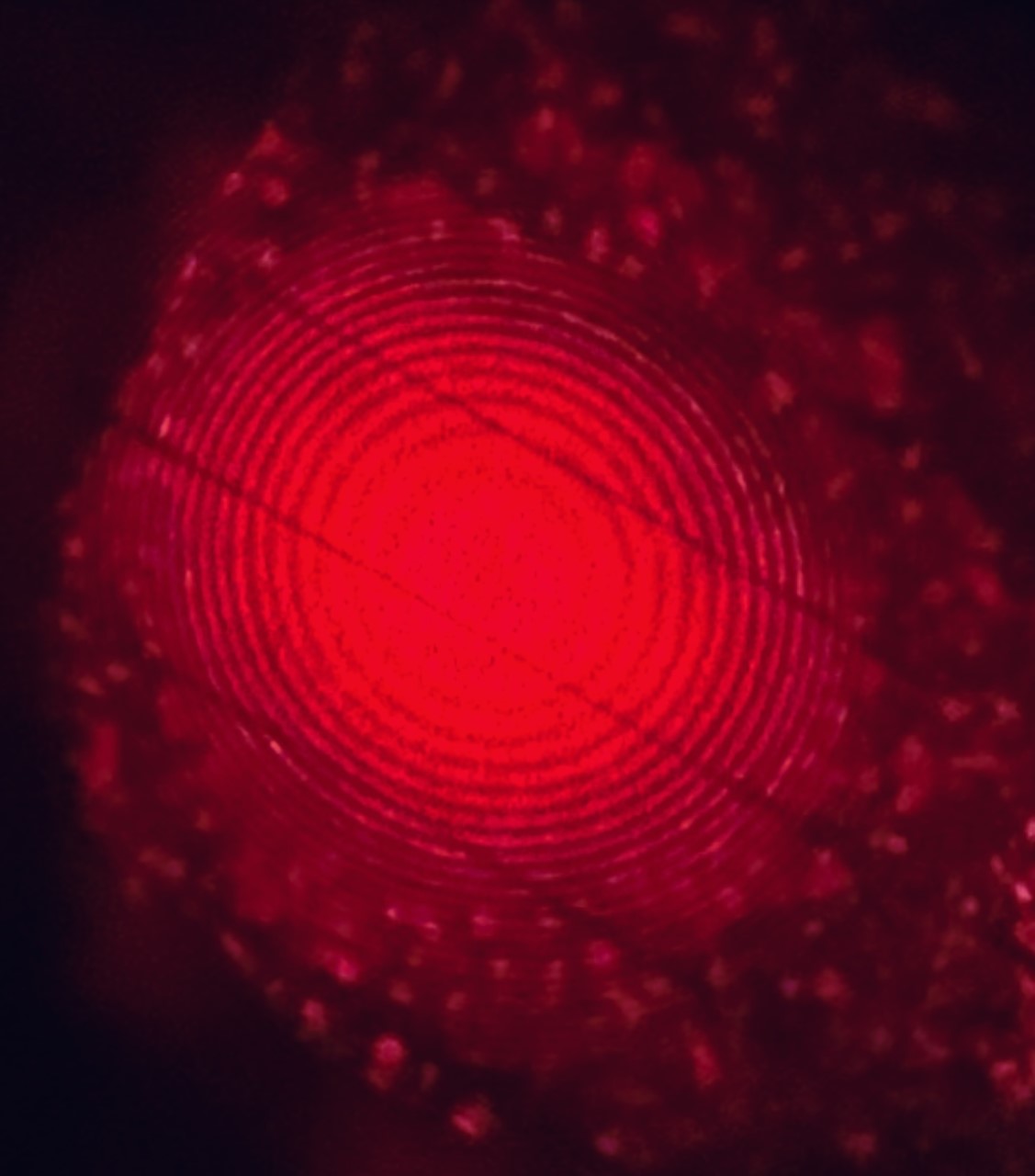}
\caption{Comparison of double fringe patterns obtained with glass slide (left) versus exemplary interference pattern obtained with silicon wafer (right) in the same experimental conditions.}
\label{fig:firstnice}
\end{figure}

\begin{figure}[h]
\centering
\includegraphics[height=1.5in, width=1.5in]{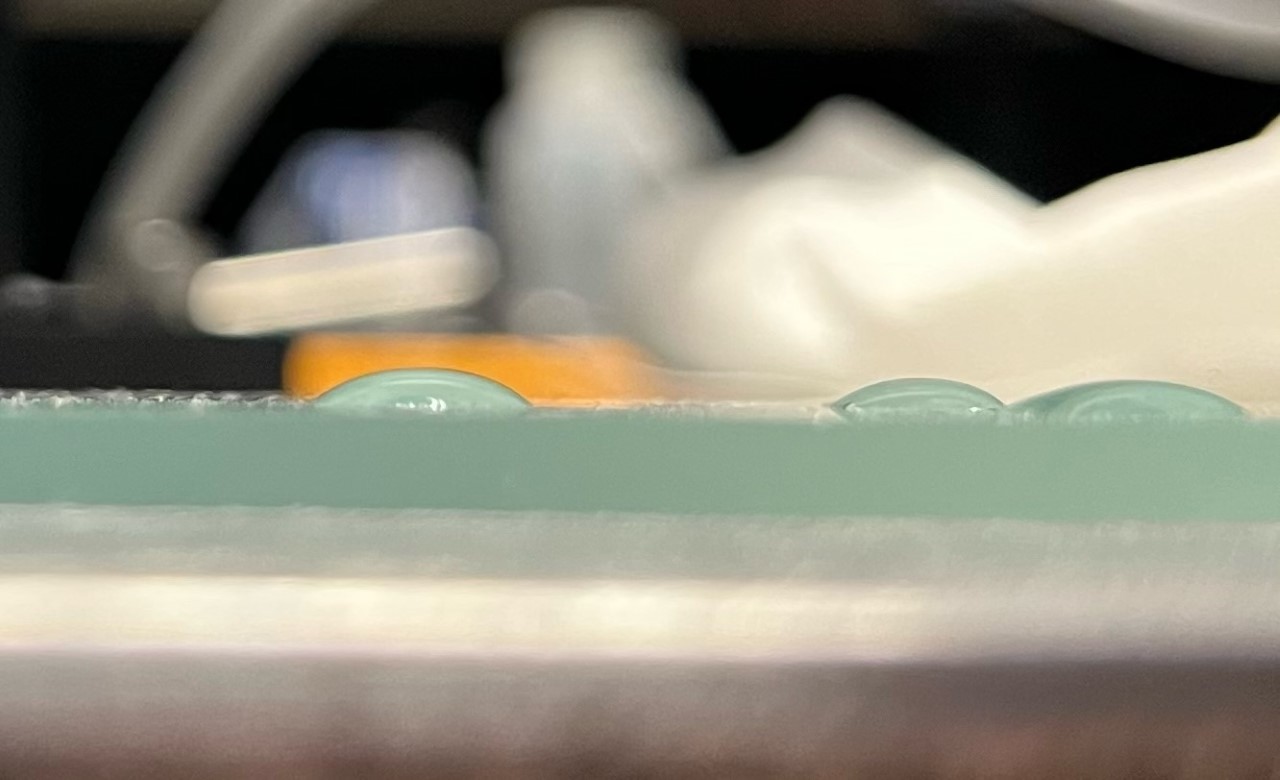}
\includegraphics[height=1.5in, width=1.5in]{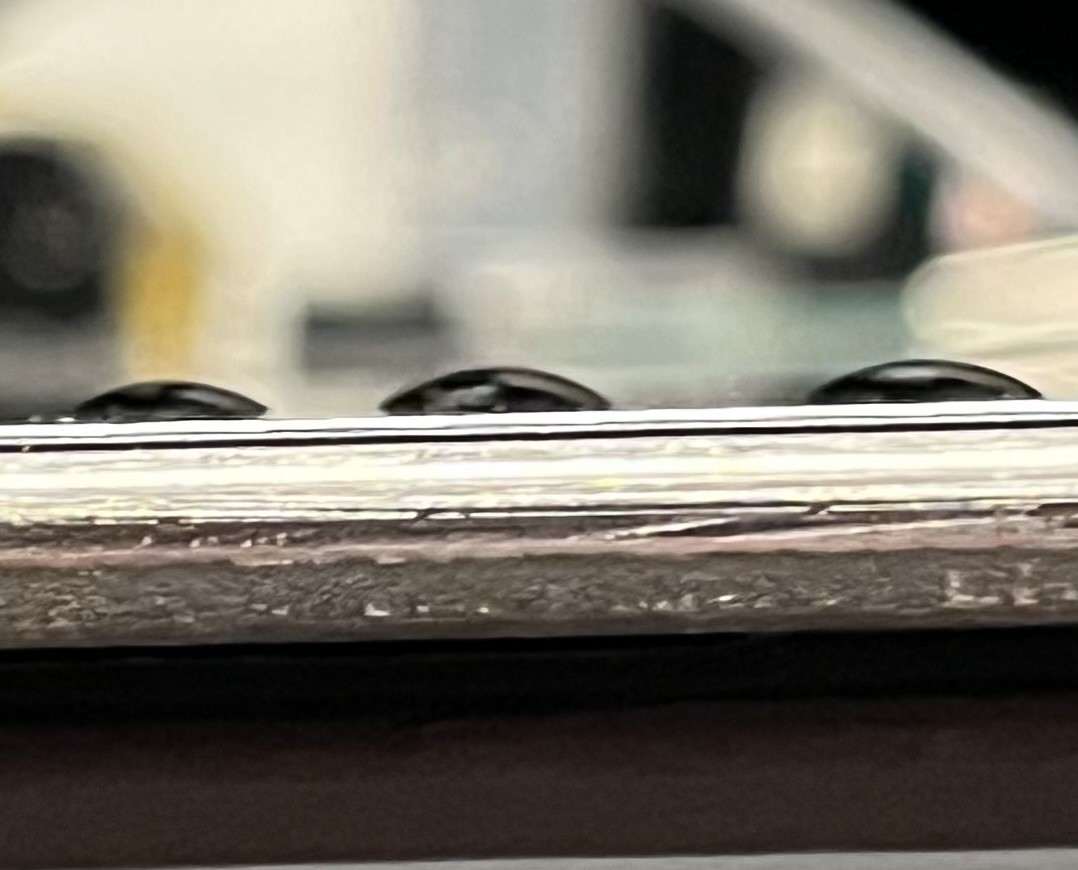}
\caption{Profile pictures of water droplet shapes on a glass slide (left) versus a silicon wafer (right). The standard droplet size is the leftmost droplet on the glass, and the middle droplet on the silicon wafer.}
\label{fig:dropletsilicon}
\end{figure}

\noindent
While the replacement of the squirt bottle with a pipette reduces the need to adjust the alignment of our optical system, another issue is the presence of two superimposed dynamic interference patterns on the observation screen. More specifically, one pattern is smaller, brighter and moves more slowly than the other. While both patterns receive the light initially reflected from the droplet surface, it is hypothesized that each pattern receives light from a different glass interface. In other words, one pattern receives light from reflection off of the water-glass interface, while the other receives light from the glass-optical table interface. The glass-optical table interface produces a larger and more diffuse fringe pattern because light becomes trapped and scatters more in the glass slide instead of going directly up to the observation screen. On top of being larger and more diffuse, this fringe pattern also moves faster than the other pattern because the change in droplet size during evaporation affects the path length difference more strongly for rays that take more time to re-emerge from the droplet. Since these superimposed patterns do not overlap perfectly, it is likely that the two glass interfaces are not parallel. This non-uniform thickness was confirmed when a new dynamic interference pattern was obtained by simply translating the glass slide under the laser without a droplet; these fringes appear related to Fizeau fringes (fringes of equal thickness). 
\\
\\
\noindent
The substrate was switched for a silicon wafer. Switching to this new substrate completely eliminated the issue of the double interference pattern experienced with the glass slide, and dynamic interference patterns were obtained that even seemed to encode the shape of the droplet within the pattern (see FIG \ref{fig:firstnice}). However, it is important to consider that since the droplet acts as a converging lens for the rays reflecting off of the substrate, the equilibrium surface tension and resulting curvature of the droplet are relevant parameters for the characteristics of our imaged pattern. Switching substrate materials may significantly alter the shape of the water droplet, but as seen in FIG \ref{fig:dropletsilicon}, the droplet shape is quite similar between these two substrates. 
\\

\begin{figure}[h]
\centering
    \includegraphics[height=1.5in, width=1.5in]{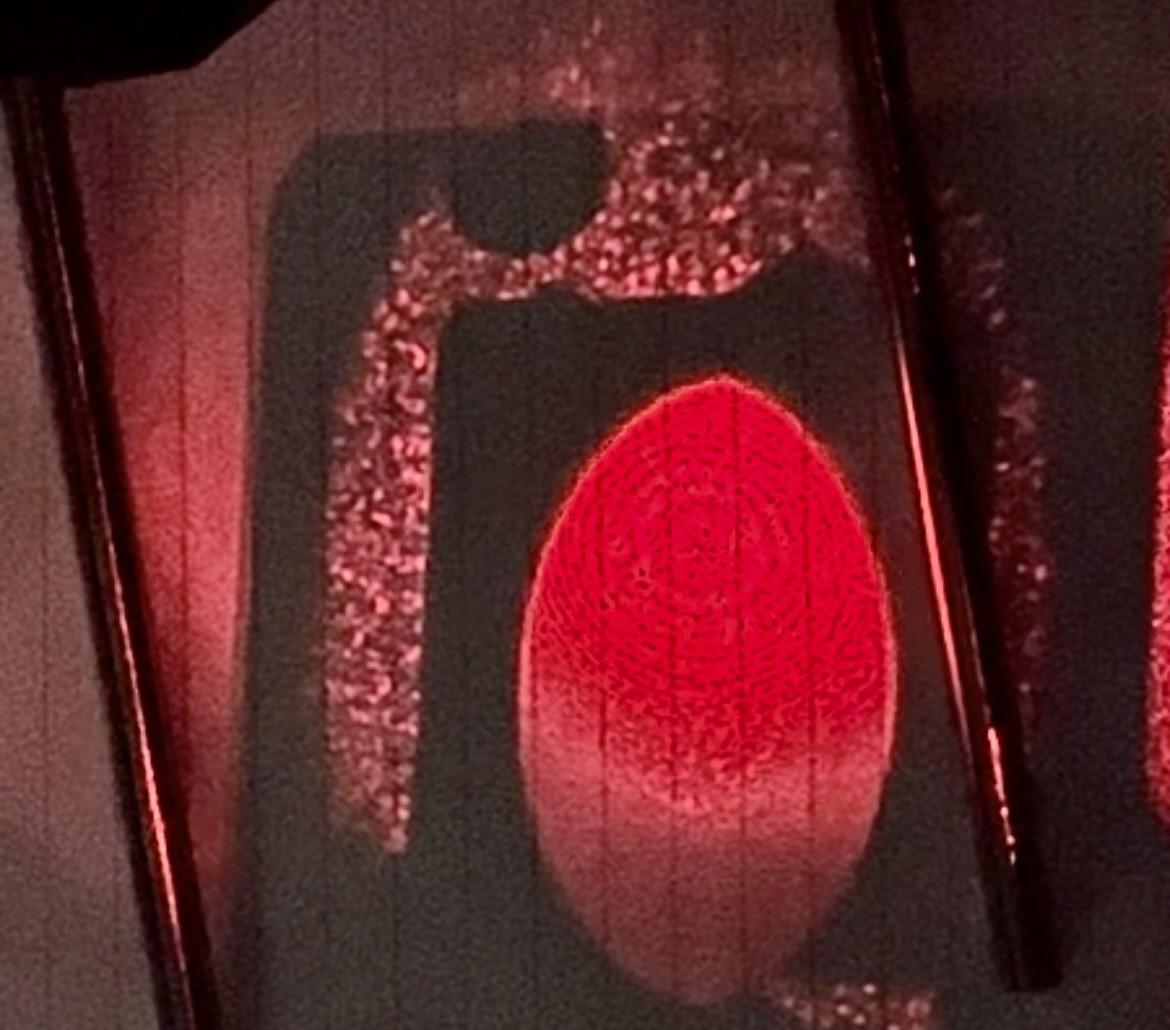}
\hspace{0.25in}
    \includegraphics[height=1.5in, width=1.5in]{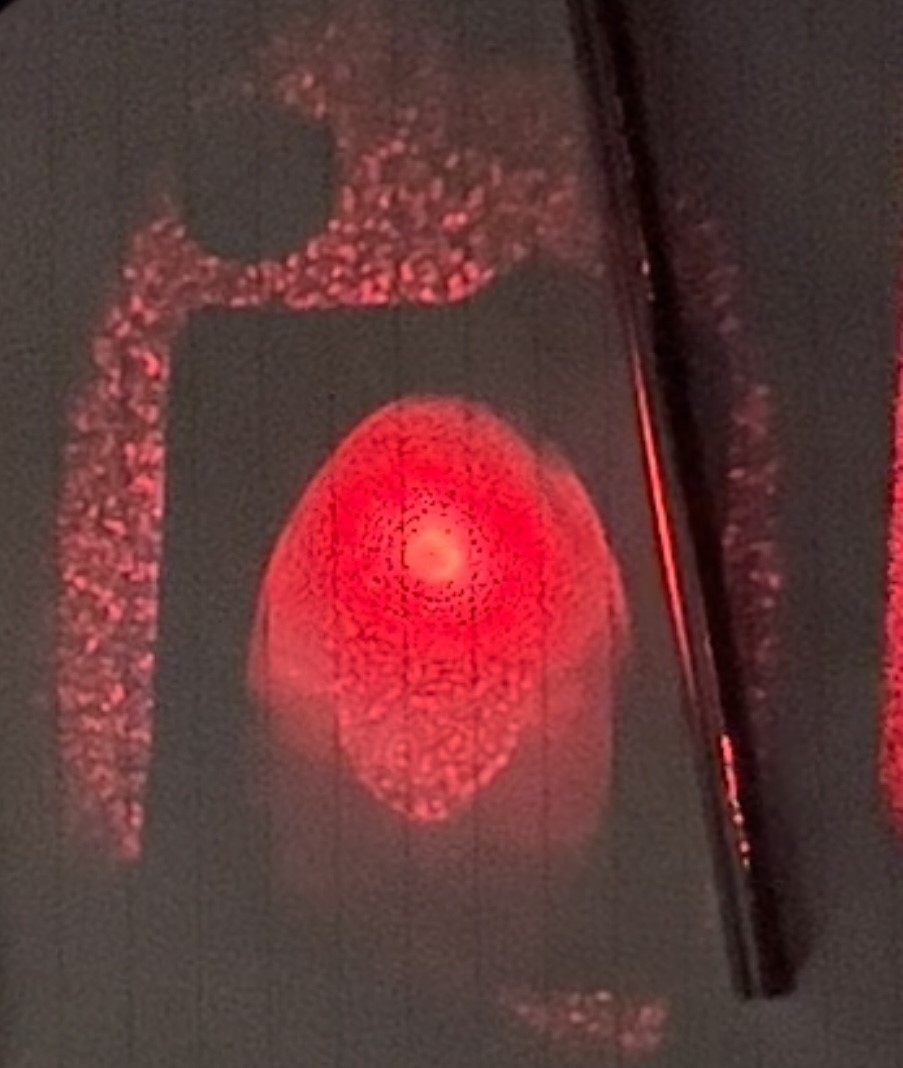}
\par\bigskip
    \includegraphics[height=1.5in, width=1.5in]{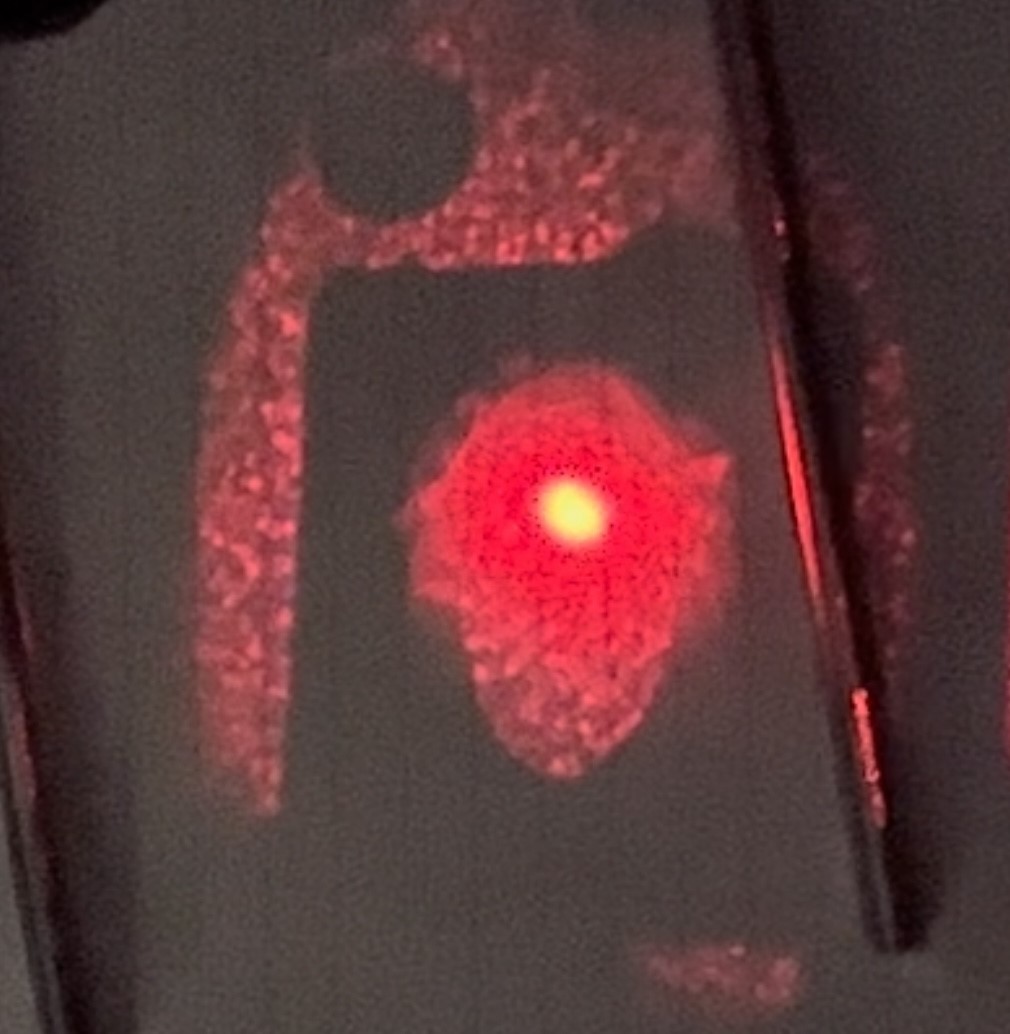}
\hspace{0.25in}
    \includegraphics[height=1.5in, width=1.5in]{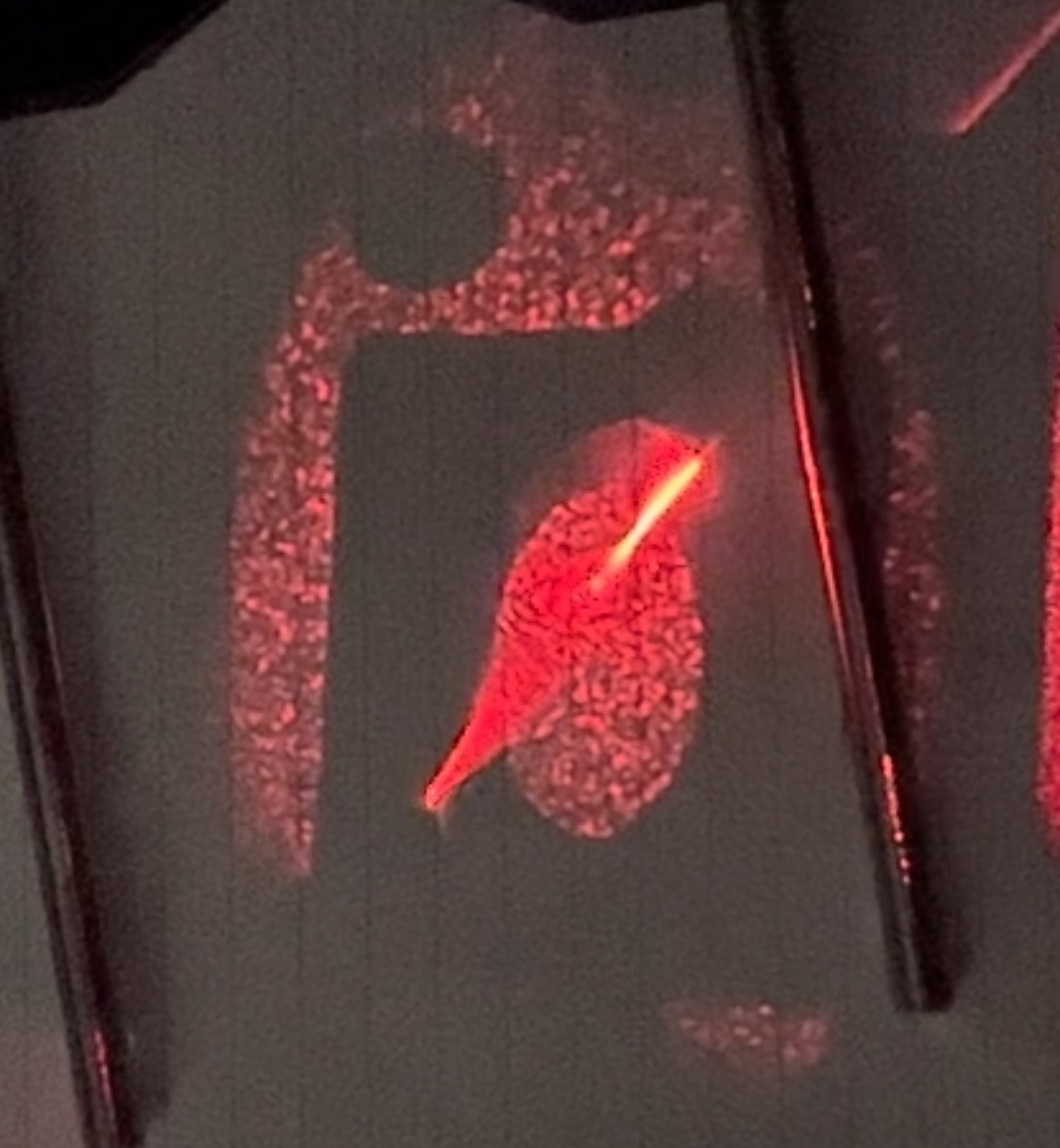}
\caption{Four low-resolution moments during the 15-minute timelapse video. The laser light gradually gets brighter at the center of the pattern, until the collapse of the droplet occurs (bottom right). The black shadow above the beam splitter shadow comes from another water droplet on the silicon wafer.}
\label{fig:timelapse}
\end{figure}

\noindent
Now, a 15-minute timelapse video is taken of the fringes produced by a water droplet evaporating on a silicon wafer as an additional confirmation of the underlying cause of our dynamic fringe pattern. Footage was taken for a droplet size smaller than the standardized droplet produced by the pipette so that the evaporation process would finish faster; the standard droplet size takes closer to an hour to fully evaporate under our experimental conditions. It was observed that the rate of fringe passage very slowly decreased over time, hinting at changes in droplet dimension during the evaporation process. The spread of reflected light imaged on the observation screen converged to a point as the droplet underwent evaporation, demonstrating the effect of a change in droplet curvature on its lens-like properties. As the droplet evaporated, its curvature seemed to decrease, which in turn would increase its associated focal length. If the focal length was initially below the observation screen, the evaporation of the water droplet should make the light converge to a point on the screen as time goes on. While everything described here is shown in FIG \ref{fig:timelapse} (in low resolution), the light never truly focused to a point on our screen. The last timelapse photo depicts the moment where the water droplet collapses out of its elliptical shape. At this moment, the droplet can be seen to occupy a small fraction of its original size, and it has lost its circular shape to take on a shape closer to that of the partially collapsed droplets in FIG \ref{fig:glassdroplet}. The observations extracted from this timelapse corroborate the notion that dynamic fringe patterns are arising from the spontaneous evaporation of the water droplet.

\noindent
At this point, it can be confidently asserted that microscopic changes in droplet size are producing macroscopic changes in the imaged interference pattern. The highly reflective silicon wafer eliminated the second dynamic pattern created by the glass slide, and the static pattern assumed to be a case of Newton's rings from the beam splitter surface is no longer prominent on the observation screen. However, Newton's rings were still visible if the observation screen were to be moved laterally to image more divergent rays; this static pattern was therefore not eliminated entirely, but it conveniently disappeared from the portion of the screen containing our desired interference pattern, hinting at a fortunate change in alignment. Despite the progress made so far in imaging reflections from an evaporating water droplet, the problem of vibrations within the dynamic interference pattern became the primary focus of the optimization process. 
\\
\\
It was found that blowing lightly on the droplet and tapping the optical table near the location of the silicon wafer induced disruptions in the dynamic fringe pattern that were accentuated versions of the vibrations previously identified. However, these transient vibrations occurred even when motion and talking ceased in the lab environment, leading us to hypothesize that it is rather the energetics associated with the three-phase boundary (air, water, substrate) that is causing these vibrations. This is supported by the observation that when two droplets were placed side-by-side on the substrate, the independently imaged interference patterns did not generally possess the same degree of vibration; these two droplets differed by their placement on the substrate and by small variations in dimension. The influence of the substrate-droplet interface has already been demonstrated in FIG \ref{fig:glassdroplet}, with the partial collapse of water droplets due to the surface characteristics of the glass slide. 
\\
\\ 
The simplest case of surface characteristics affecting droplet shape is the cleanliness of the substrate. It is very important to avoid the accumulation of dust and fingerprints on the substrate, as these particles can get sucked into the water droplet and float around on its surface, thereby contaminating the resulting interference pattern. Early on, wax had been placed on a glass slide in contact with the hot plate. The applied heat caused pieces of wax to detach from the glass and float across the droplet, causing the formation of mobile shadows on the observation screen. Once the wax and hot plate were abandoned, this type of contamination was still present; when the water droplets fully evaporated, a circular trace of contaminants was left behind right were the droplet perimeter used to be. Until now, tap water was being used to create the droplets, but a switch was made to distilled water in the event that minerals or other contaminants were present in significant amount in the tap water. This reduced the trail left behind by the evaporated droplets, but it did not eliminate the trail entirely. 
\\
\\
It is evident that the microscopic structure of the underlying substrate influences the imaged interference pattern. In order to further probe these influences, four different substrates were tested to verify their ability to produce a clear dynamic fringe pattern: a glass slide made of fused silica, a glass slide made of fused silica with a coating of Howie's hockey stick wax, a silicon wafer, and a silicon wafer that had undergone a buffered-oxide etch, thereby leaving a hydrogen-terminated silicon surface. A profile picture of the standardized water droplet is taken for each of these substrates to visualize the contact angles formed at the three-phase boundary, allowing for comparisons to be made with the simulated droplet geometry. Each of these are then illuminated by laser light to identify which substrate is the most suitable for obtaining a clear interference pattern.

\section{Results and Discussion}

\subsection{Simulation vs. Experiment}

\begin{figure}[H] 
    \centerline{\includegraphics[scale=0.19]{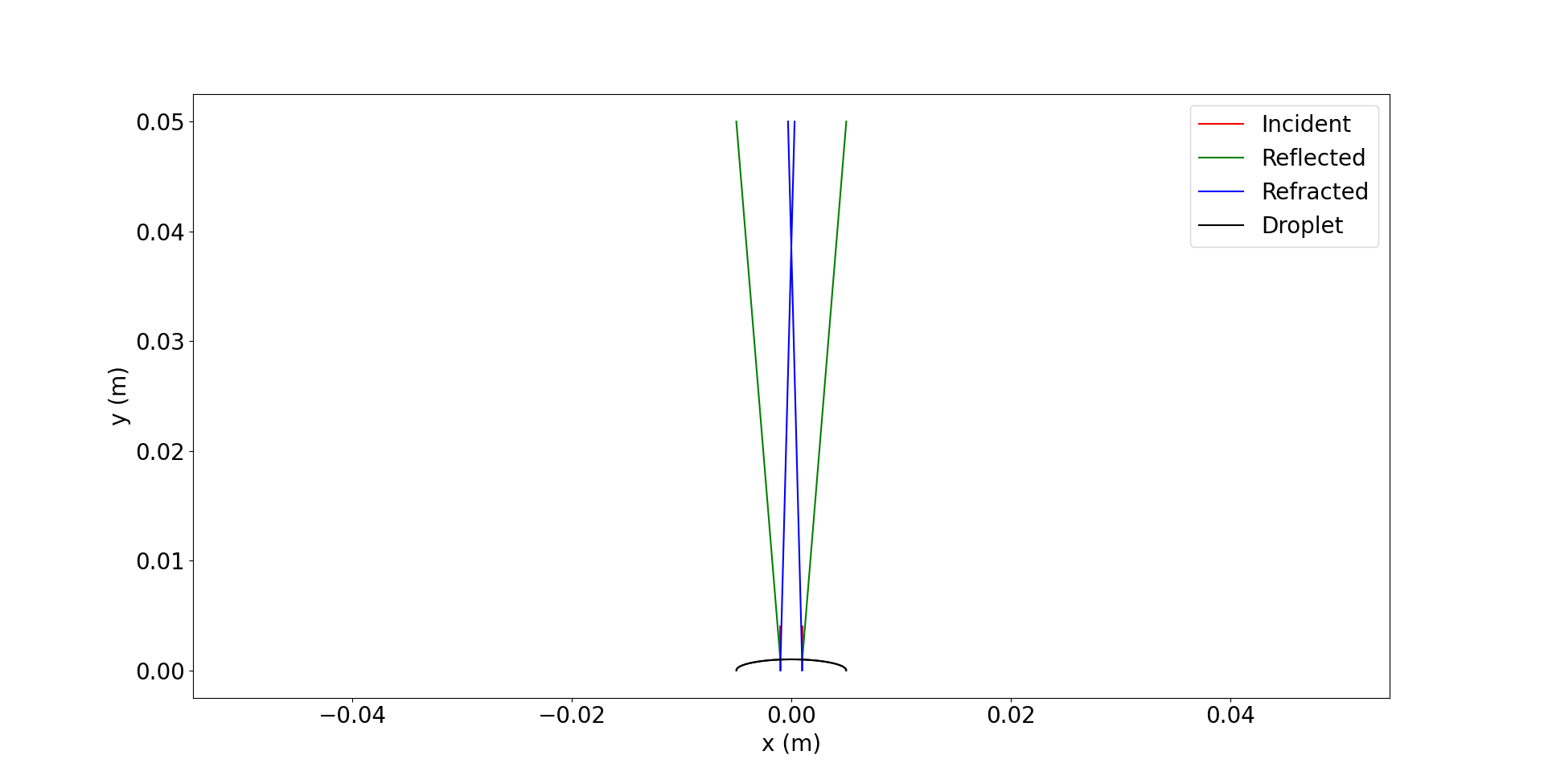}}
    \caption{Low ray density graph depicting the focal point of our modelled water droplet. The droplet is modelled with a refractive index of $n=4/3$, a semi-major radius of 5 mm and a semi-minor radius of 1 mm. Green rays depict reflected rays, while blue rays represent refracted rays.}
    \label{fig:FocusPoint}
\end{figure}

\noindent
As alluded to previously, the droplet acts as a converging lens for the rays undergoing refraction. For a droplet with a width of 5 mm and a height of 1 mm, the simulation predicts the focal point to be located approximately 4 cm above the droplet, as shown in FIG \ref{fig:FocusPoint}. G-L. Ngo et al.'s findings suggest that the direction of fringe passage changes as the observation screen passes through the focal point\cite{template}. Note that this could not be confirmed experimentally due to limitations in our setup; the height of the beam splitter was around 15 cm above the droplet, so the screen could not be placed near the focal point. However, our simulation produced results confirming that the fringes move outwards when the screen is above the focal point, and the fringes move inwards when the screen is below the focal point; the interference pattern is not present at the focal point itself, as all information about the changing droplet shape that is contained in the refracted rays is lost when it collapses to a point.

\noindent
The following simulation results were produced by splitting the incident laser light into 3000 rays. The refractive index of air is taken to be 1, and the refractive index of water as well as the droplet dimensions are the same that were used to generate FIG \ref{fig:FocusPoint}. The screen was placed at a distance of 4 mm from the droplet, and so this is a screen height where the dynamic fringe pattern collapses inward with time. While the direction of fringe passage is qualitatively different at this screen height than at the screen height used experimentally ($\approx50 cm$), many comparisons can still be made between both results; note that the horizontal scale on simulated interference patterns will be specific to the chosen height of the observation screen, and so if one seeks to use the spacing between fringes or the rate of fringe passage to extract information about changes in the optical path length, both experimental and simulated screen heights must match. 
\\
\\
We used three different models to consider the intensity reaching the observation screen due to the reflected and refracted rays: equal intensity in reflected and transmitted rays, Schlick's approximation, and the Fresnel model. First, we will focus on the difference between the results produced by the Schlick approximation and by the Fresnel model to assess the validity of the former.

\begin{figure}[H] 
    \centerline{\includegraphics[scale=0.19]{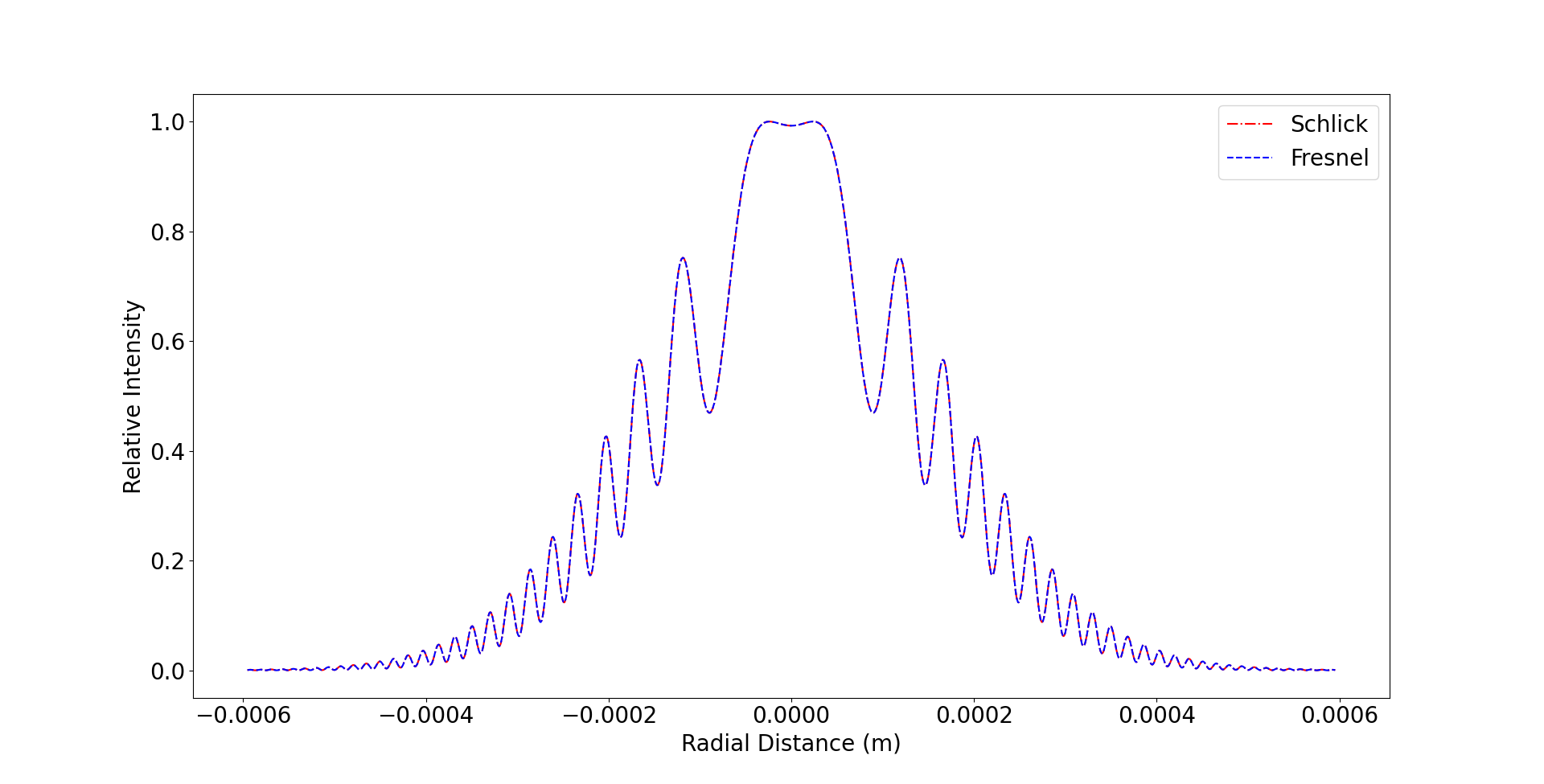}}
    \caption{The intensity reaching the observation screen as a function of the radial distance is plotted for both the Schlick (red) and Fresnel (blue) models. The overlap between both models makes the plot look purple.}
    \label{fig:Fresnel_2d}
\end{figure}

\begin{figure}[H] 
    \centerline{\includegraphics[scale=0.19]{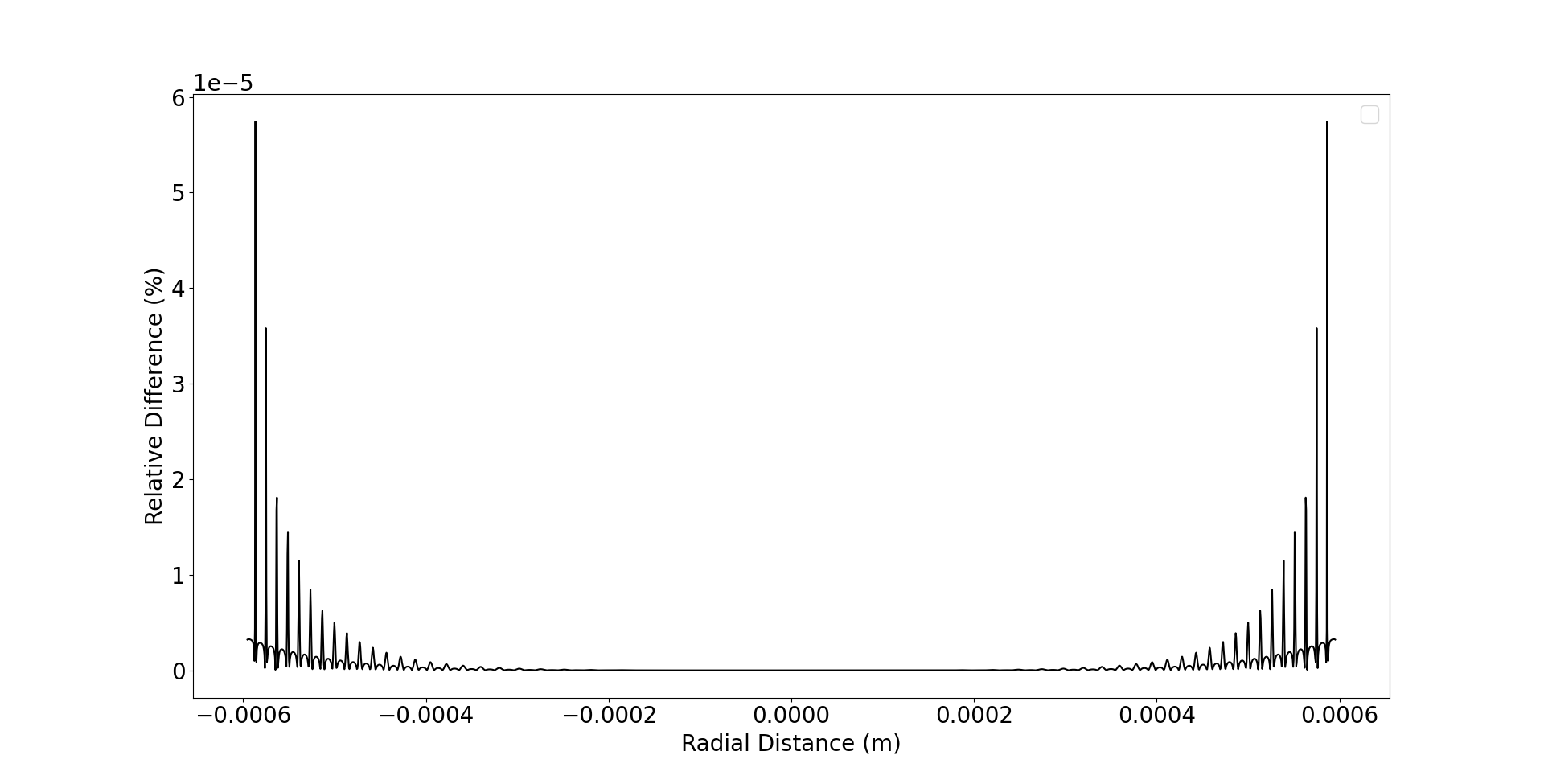}}
    \caption{Relative difference between the Fresnel and Schlick models as a function of radial distance from the center of the screen.}
    \label{fig:Relative_difference}
\end{figure}

\begin{figure}[H] 
    \centerline{\includegraphics[scale=0.19]{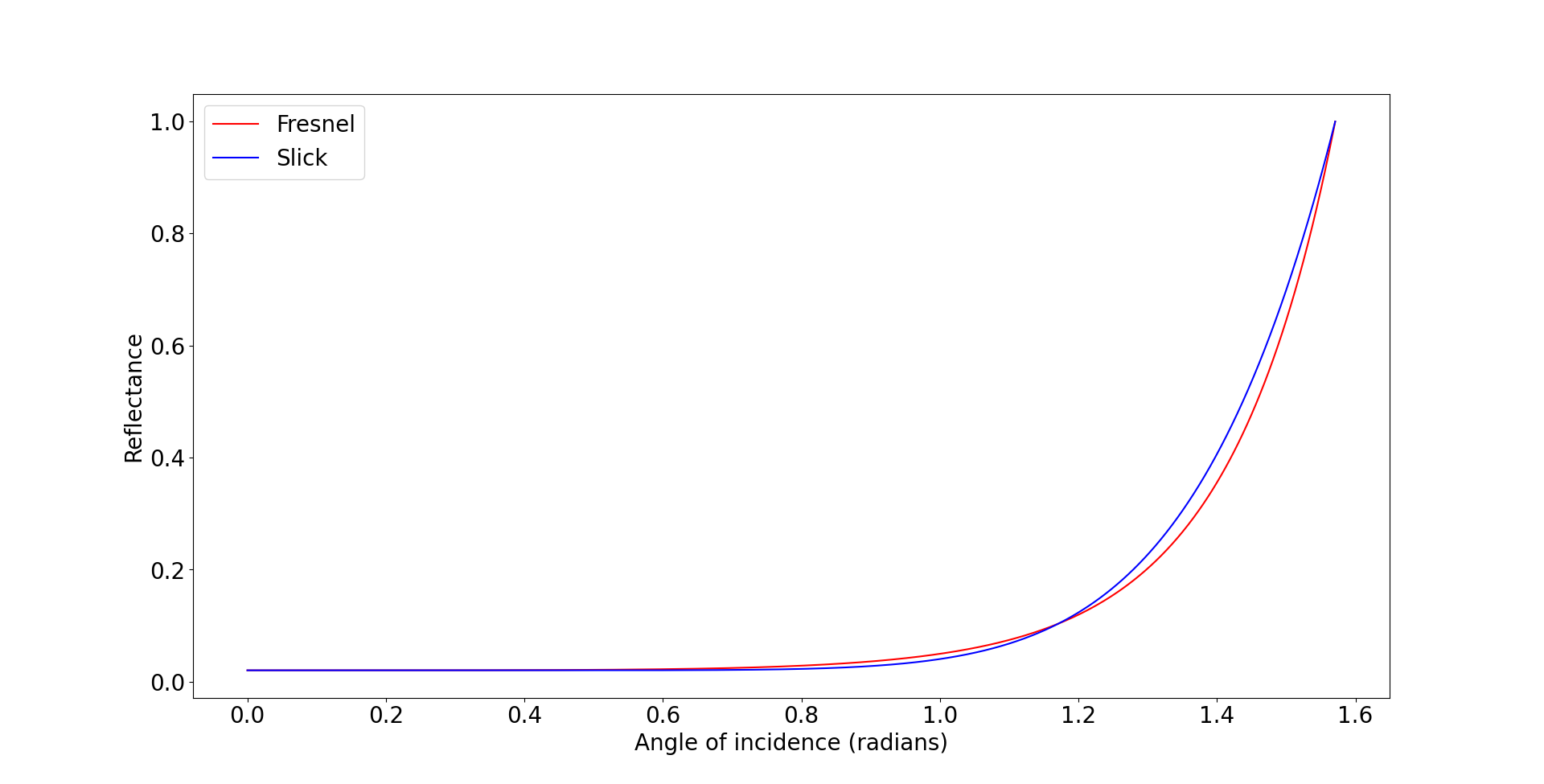}}
    \caption{Reflectance as a function of angle of incidence for the Schlick and Fresnel models. The entire range of incident angles are plotted to show the utility of Schlick's approximation.}
    \label{fig:Fresnel_Slick}
\end{figure}

\noindent
The complete overlap in FIG \ref{fig:Fresnel_2d} shows that there is practically no difference between the intensity distributions given by the Schlick and Fresnel intensity models for our system. The relative difference between these two models is calculated and plotted in FIG \ref{fig:Relative_difference}. As a confirmation, the relative difference graph shows that Schlick's approximation matches the results derived from Fresnel's equations for small radial distances. FIG \ref{fig:Fresnel_Slick} further illustrates the congruence between both models, showing their predicted reflectances as a function of angle of incidence. At larger radial distances, the two models deviate, but this corresponds to spots on the droplet surface for which reflected rays diverge too strongly to get imaged on the observation screen; in particular, the discrepancies in FIG \ref{fig:Relative_difference} correspond to approximately the 0.4 to 1 radians region on the graph in Figure \ref{fig:Fresnel_Slick}, and these are incident angles that diverge considerably, depending on the droplet curvature. Thus, Schlick's approximation is indeed computationally-efficient as it produces the same results as that derived from Fresnel's equations, but with a simpler mathematical approach. We will now compare this approximation with a model that assumes light intensity is split equally between the transmitted and reflected rays.

\begin{figure}[H] 
    \centerline{\includegraphics[scale=0.19]{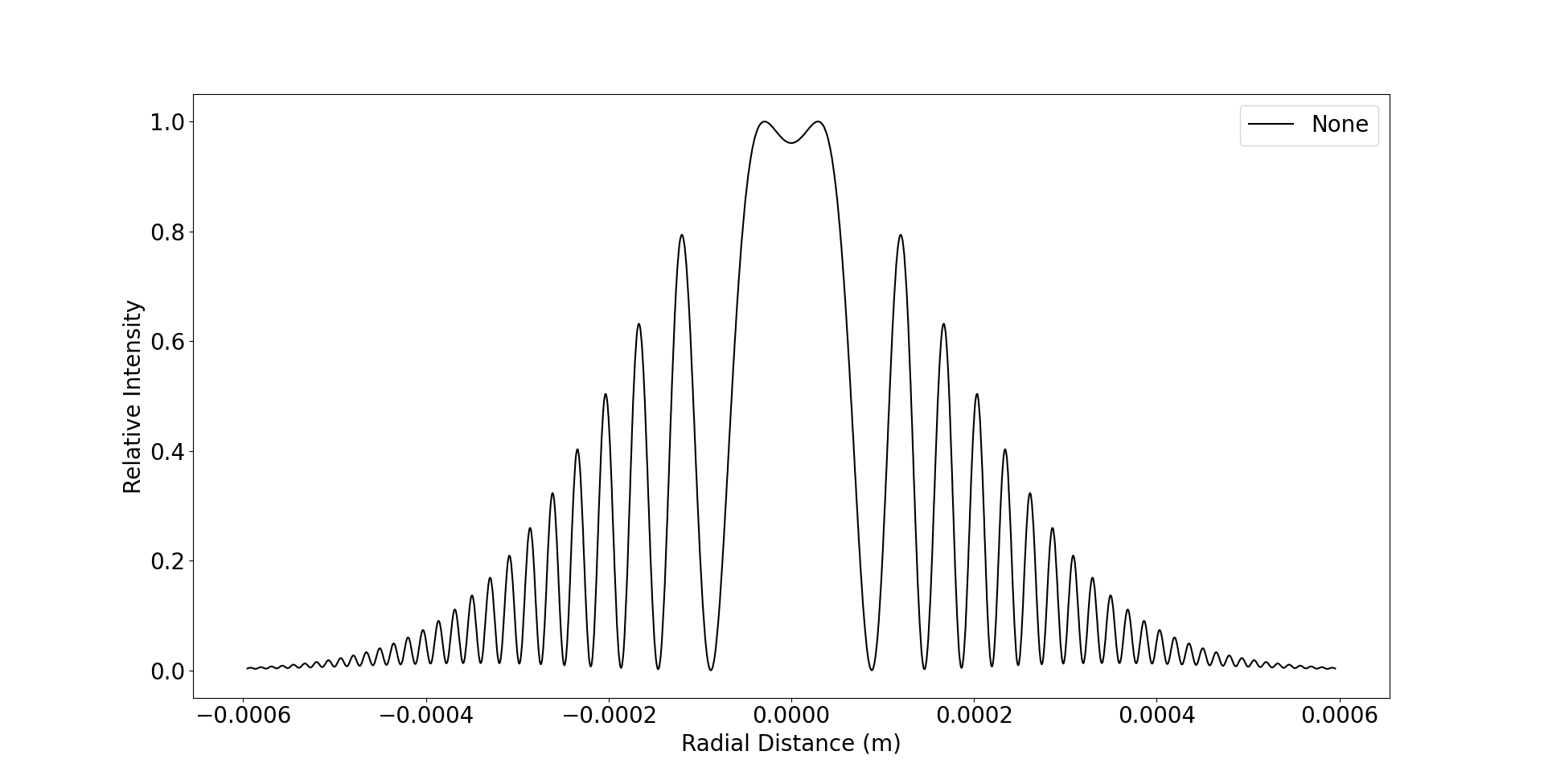}}
    \caption{Relative intensity of interference as a function of radial distance from the center of the screen for equal intensity between reflected and transmitted rays. All dark fringes possess the same intensity in this model.}
    \label{fig:Fresnel_nothing_close}
\end{figure}

\begin{figure}[H] 
    \centerline{\includegraphics[scale=0.19]{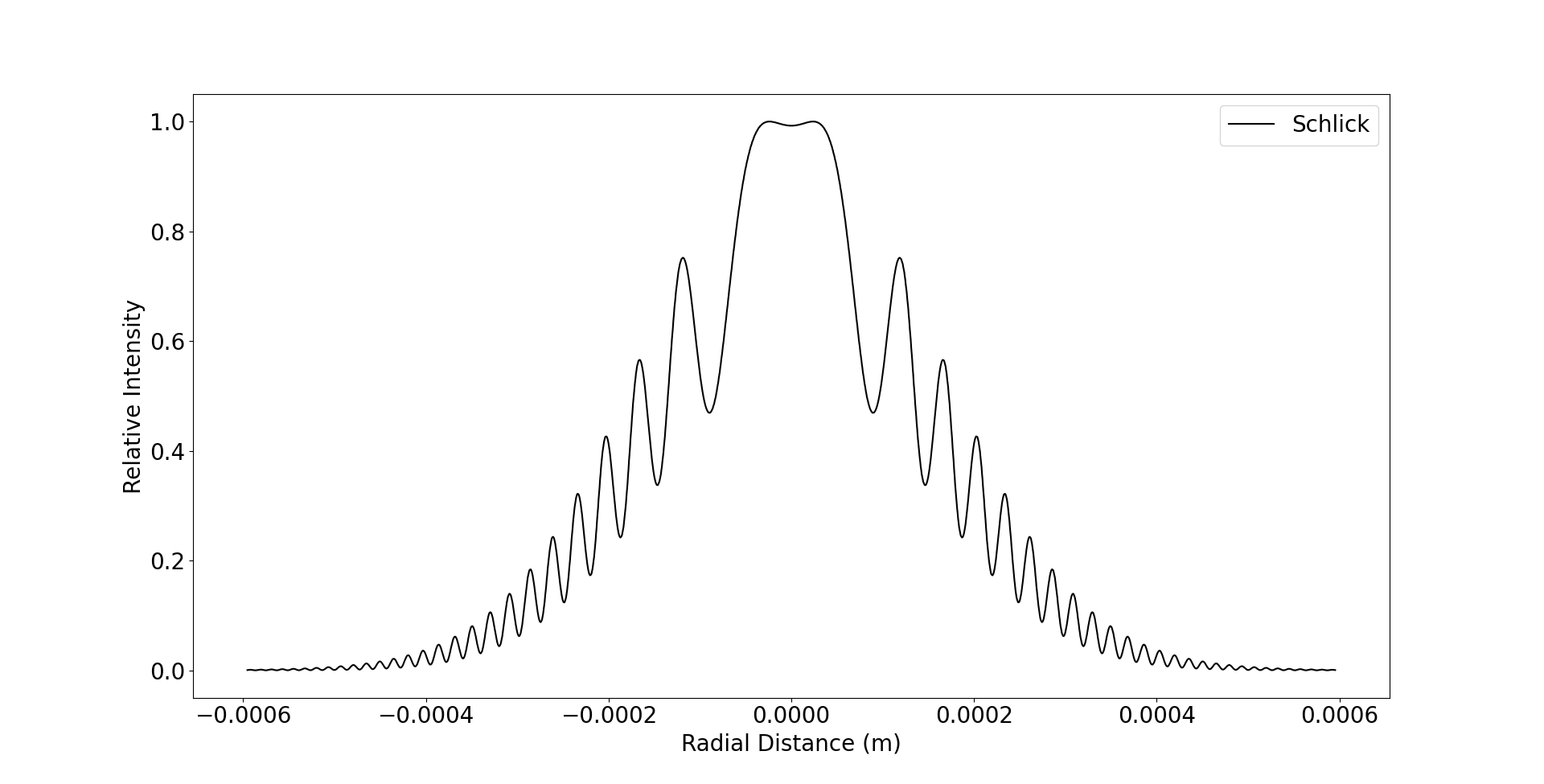}}
    \caption{Relative intensity of interference as a function of radial distance from the center of the screen using the Schlick model. The intensity of the dark fringes decays as a function of radial distance.}
    \label{fig:Fresnel_schlick_close}
\end{figure}

\begin{figure}[H] 
    \centerline{\includegraphics[scale=0.19]{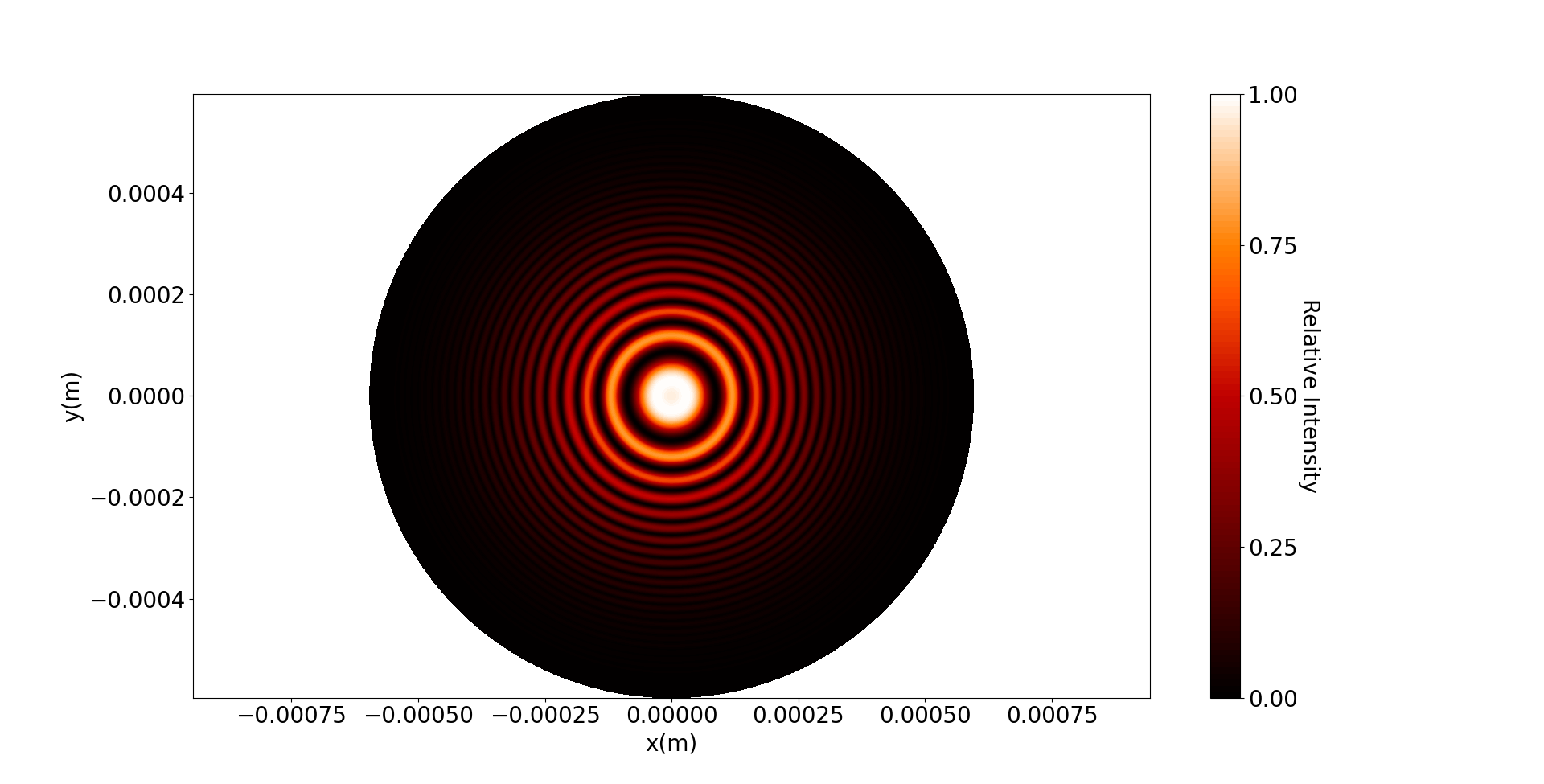}}
    \caption{Contour plot of the relative intensity as a function of radial distance from the center of the screen for equal intensity between reflected and transmitted rays.}
    \label{fig:Fresnel_3d_none}
\end{figure}

\begin{figure}[H] 
    \centerline{\includegraphics[scale=0.19]{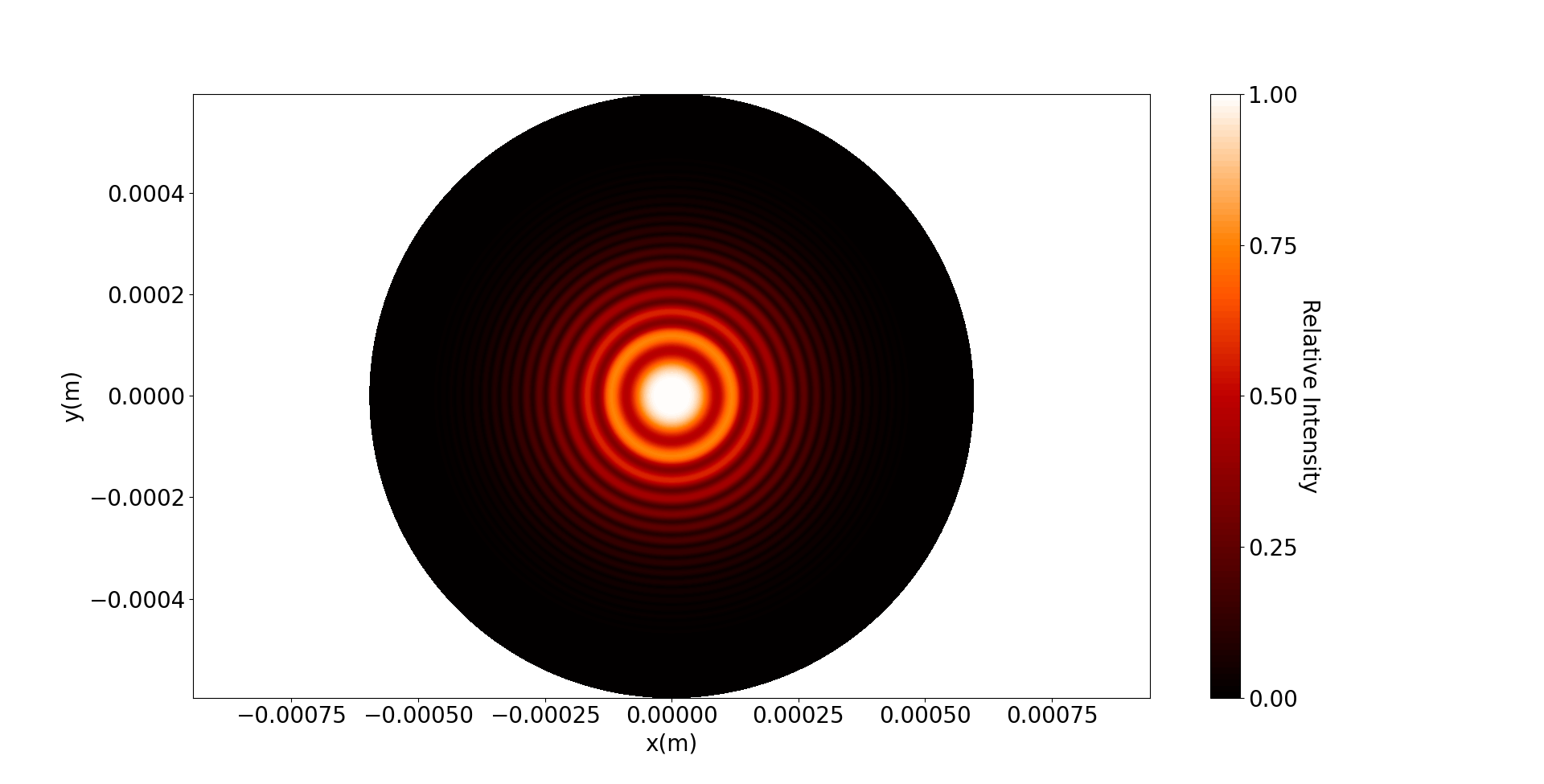}}
    \caption{Contour plot of the relative intensity as a function of radial distance from the center of the screen using the Schlick approximation.}
    \label{fig:Fresnel_3d_schlick}
\end{figure}

\noindent
When comparing FIG \ref{fig:Fresnel_nothing_close} and \ref{fig:Fresnel_schlick_close}, it can be seen that Schlick's approximation causes the bright fringe intensity to drop off slightly faster with radial distance compared to when light is split equally between reflected and transmitted rays. It can also be seen, demonstrated further in FIG \ref{fig:Fresnel_3d_none}, that the dark fringes all possess the same minimum intensity if Schlick's approximation is not applied. In contrast, FIG \ref{fig:Fresnel_schlick_close} and \ref{fig:Fresnel_3d_schlick} demonstrate that the dark fringes do not all possess the same intensity when using Schlick's approximation. In particular, the intensity of the dark fringes decreases with radial distance, along with the bright fringes. Therefore, Schlick's approximation makes for less contrast between adjacent bright and dark fringes, as shown by the lower fringe resolution in FIG \ref{fig:Fresnel_3d_schlick} as compared to FIG \ref{fig:Fresnel_3d_none}. An interesting consequence of Schlick's approximation is that dark fringes near the center may appear brighter than some of the higher order bright fringes.

\begin{figure}[H] 
    \centerline{\includegraphics[scale=0.30 , angle =90]{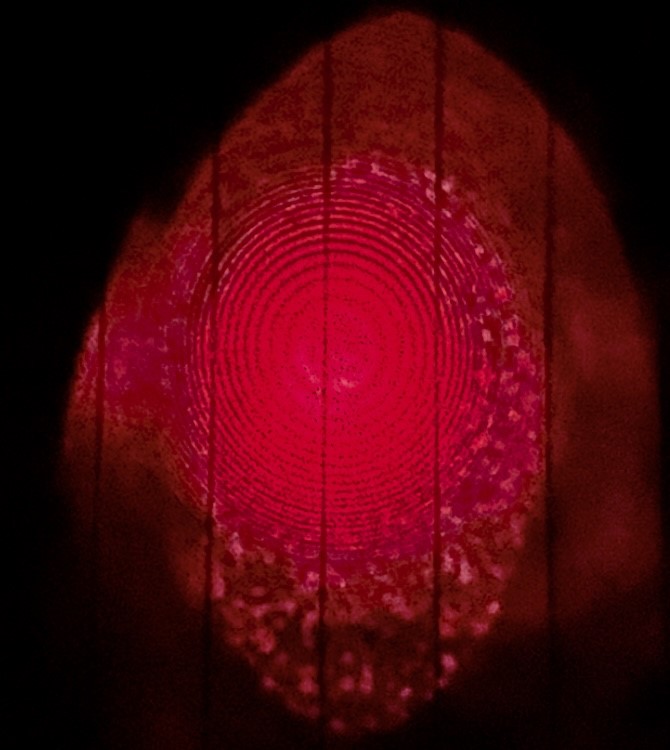}}
    \caption{Experimental interference pattern obtained using a silicon wafer as a substrate, with the observation screen placed at about 50 cm above the droplet.}
    \label{fig:beautypattern}
\end{figure}

\noindent
When comparing the previous observations to a real image of our interference pattern, as shown in FIG \ref{fig:beautypattern}, we found that  Schlick's approximation yields more physically accurate results. The sharp contrast between each bright and dark fringe predicted in FIG \ref{fig:Fresnel_3d_none} is not present in the real pattern. The real pattern's central dark fringes possess a similar intensity to the next order bright fringes, making for a lower fringe resolution in the center of the pattern, as seen in FIG \ref{fig:Fresnel_3d_schlick}. Therefore, it is important that our model correctly accounts for light loss at each optical boundary in order to simulate a realistic interference pattern. However, our model does not account for the influence of multiple reflections in the droplet eventually reaching the screen. We expect that these ignored interactions, as well as scattering throughout our optical system (impurities on wafer surface, in the water, or on beam splitter) gives rise to the more diffuse light reaching the screen in FIG \ref{fig:beautypattern}.
\\
\\
Two models for the change in droplet dimension during evaporation are presented, and Schlick's approximation is used to assess the intensity distribution reaching the observation screen. To compare the models, we recorded a 2000-frame video at a rate of 50 frames per second, using a screen positioned 5 mm above the droplet. The starting values for $R_x$ and $R_y$ were 5 mm and 1 mm, respectively. For the first model, we decreased both $R_x$ and $R_y$ at a constant rate of $\num{4e-7}$ mm per frame. For the second model, we maintained the same rate but only changed $R_y$. Videos depicting the results of each model can be found in Appendix B.
\\
\\
With the initial droplet geometry and evaporation rate that we used, there was barely any difference between the two evaporation models over the time interval considered. Both models treat $R_y$ identically, but they differ in their treatment of $R_x$. Since shrinking both dimensions produced a nearly identical image as just shrinking the height did, it can be concluded that while shrinking $R_x$ does change the droplet curvature, which would change the rate of fringe of passage observed on the screen over time, changing the droplet height ($R_y$) more directly affects the optical path length of the refracted rays. Also, note that it is unfavourable for the droplet to shrink too quickly in $R_x$, because this would increase the droplet curvature, thereby raising the interfacial stress experienced by the droplet. In reality, evaporation in a constant temperature and pressure environment is likely a hybrid of these two processes, with the exact rates of dimension change depending on the three-phase boundary under consideration as well as the initial droplet dimensions.

\subsection{Experimental Substrate Comparison}

\begin{figure}[h]
\centering
\includegraphics[height=1.5in, width=1.5in]{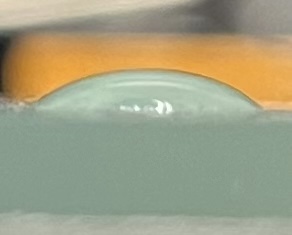}
\includegraphics[height=1.5in, width=1.5in]{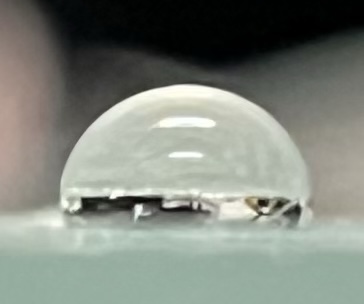}
\caption{Comparison of droplet contact angle on glass-based substrates. On the left is a plain glass slide, and on the right is a glass slide with a coating of Howie's hockey stick wax.}
\label{fig:glassshape}
\end{figure}

\noindent
After having completed the setup optimization process, a droplet of similar volume was placed on four different substrates for comparison. An ellipsoidal droplet shape is desired to match the geometry used in the simulation. Moreover, the substrate must reflect light back towards the observation screen in such a way that a clear interference pattern is imaged.
\\
\\
FIG \ref{fig:glassshape} shows the two glass-based substrates. The plain glass slide produces a droplet shape with a curvature that gives credence to our ellipsoidal model. This figure also displays the resulting droplet shape after applying a uniform layer of Howie's hockey stick wax. This second droplet has a much greater curvature, typical of a more hydrophobic surface, but this enhanced curvature visibly exceeds that used in the model, resembling more a sphere than an ellipsoid. The wax layer, despite being relatively uniform, appears to scatter light too strongly for a significant amount of reflected light to make it up the observation screen. The wax-coated glass slide is therefore not a good choice in substrate for producing a clear dynamic fringe pattern. While the uncoated glass slide does manage to produce an interference pattern, FIG \ref{fig:firstnice} demonstrates its tendency to form two distinct overlapping interference patterns caused by reflections from each glass boundary. This double reflection should be avoided for a clearly imaged pattern, and the potential for non-uniform thickness worsens the glass slide's ability to serve as a substrate. However, our interferometer's ability to detect microscopic changes in droplet dimension can still be confirmed with this otherwise suboptimal substrate. 
\\

\begin{figure}[h]
\centering
\includegraphics[height=1.5in, width=1.5in]{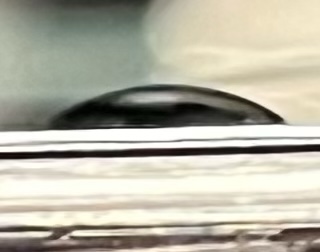}
\includegraphics[height=1.5in, width=1.5in]{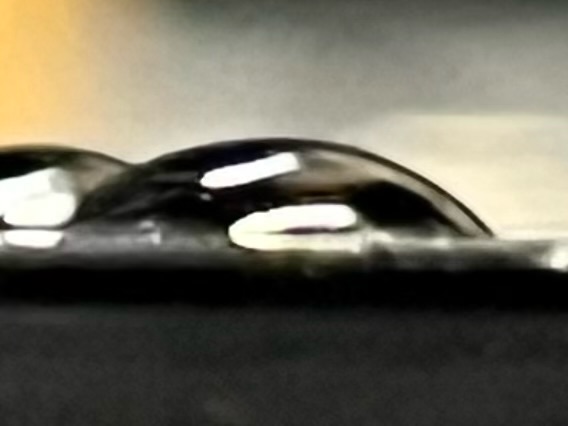}
\caption{Comparison of droplet contact angle on silicon wafer substrates. On the left is a plain silicon wafer, and on the right is a hydrogen-terminated silicon wafer surface.}
\label{fig:silishape}
\end{figure}

\noindent
As shown previously in FIG \ref{fig:dropletsilicon}, the droplet curvature does not change significantly when the substrate is switched to a silicon wafer. However, this substrate is already much better than the glass slide, as there are no more double reflections. In addition, silicon is more reflective than glass, allowing for more light to be sent back to the observation screen. The improvement in image clarity after switching to the silicon wafer from the glass slide is already depicted in FIG \ref{fig:firstnice}. It can be seen in FIG \ref{fig:silishape} that a silicon wafer having undergone a buffered-oxide etch produces a droplet with slightly more curvature than the regular silicon wafer does; this can be explained by the enhanced hydrophobicity provided by a hydrogen-terminated surface after being treated with hydrofluoric acid. While both silicon-based substrates appear to produce better interference patterns, not enough footage was taken to demonstrate the clear superiority of one over the other. However, the treated wafer did seem to produce a slightly clearer pattern than the regular silicon wafer. Its larger curvature may have focused light better onto the observation screen for our particular setup. 
\\

\noindent
While the droplet size was standardized to try to keep consistent droplet dimensions from one trial to the next, it was later noticed that this standard droplet size was usually not the droplet shape that produced the clearest pattern. Rather, the ideal droplet size was consistently a little bit smaller than the standard droplet size, and this was the case for all substrates that successfully imaged an interference pattern. Thus, after placing a droplet on the substrate it was found that a certain time (the "equilibration time") had to pass before the standard droplet evaporated into the ideal droplet size, and the equilibration time was found to differ slightly from one substrate to the next. In particular, it took about 40 minutes for the glass slide, 30 mins for the silicon wafer, and 25 mins for the treated silicon wafer. These are all rough estimates, as each measured equilibration time is attributed to the nearest multiple of 5 minutes. This equilibration time is likely linked to the transient vibrations encountered during the optimization of the setup, and it is hypothesized that it is dictated by the energetics of the particular three-phase boundary under consideration. This is supported by the change in measured equilibration time for each substrate, and it is likely no coincidence that the equilibration time decreases with increasing droplet curvature, but we did not look into this.

\section{Conclusion}

%improvements: didnt measure droplet size or assess contact angles quantitatively, didnt measure droplet volume, add lens to do focal point assessment, photodiode for better footage
\noindent
The experimental procedure outlined in this report sought out to verify that a dynamic interference pattern could be imaged as a result of the change in dimension of an evaporating water droplet. This was achieved successfully by directing Metrologic's ML868 He-Ne laser light onto a sessile water droplet of about 5 mm diameter, using a silicon wafer substrate to redirect the light to an observation screen located about 50 cm above the droplet. The resulting interference pattern was compared with results obtained through a ray optics Python simulation. The simulation predicts that an initial ellipsoidal droplet size of 5 mm wide and 1 mm tall makes for a focal length of 4 cm, above which the dynamic fringe pattern shows outward-travelling fringes as found experimentally at a screen height of 50 cm. The simulation predicts that at heights below the focal length, inward-travelling fringes will be imaged on the observation screen, while the pattern disappears at the focal length, but this was not confirmed experimentally. 
\\
\\
Three intensity models were used in simulation to compare with the intensity distribution measured experimentally. Schlick's approximation for the reflectance of a non-conducting interface produced a nearly identical intensity distribution as a model derived from Fresnel's equations for unpolarized light, due to the small angles of incidence in our optical system. Both of these models produce a more realistic interference pattern than a model that assumes equal light is sent into reflected and transmitted portions of the incident light ray. The low contrast seen experimentally between central bright and dark fringes demonstrates the importance of properly accounting for the loss of light at each optical boundary in simulation. Further, two models of droplet dimension evolution were compared, showing that for the droplet dimensions used in simulation, a slow decrease in droplet height likely has a larger effect on varying the optical path lengths in the system than does a decrease in droplet width.  
\\
\\
A brief analysis of substrate choice is performed, showing that a glass slide can give rise to double reflections that generates two dynamic fringe patterns on the observation screen, resulting in reduced visibility. The use of a hydrophobic wax coating creates a droplet with curvature exceeding that of our ellipsoid model, and it also scatters light too much to produce a clear interference pattern. The switch to silicon-based substrates produced the clearest interference patterns, and it was found that hydrogen-terminated silicon produced droplets with slightly larger curvature than an untreated silicon wafer did; a thorough comparison between the patterns imaged by each silicon substrate was not carried out. It was also found that imaging in this optical system is very sensitive to the particular three-phase boundary under study. More hydrophobic substrates produce larger curvatures, which affects the droplet dimension and induced path length differences in the optical system. An approximate time scale, named the equilibration time, is attributed to the time it takes for each substrate to allow the standard droplet size to develop into the ideal droplet size. This time is 40 minutes for a glass slide, 30 minutes for a silicon wafer, and 25 minutes for the treated silicon wafer. We suggest that this equilibration time may provide key insights into the energetics of the three-phase boundary, and may be of interest for future studies in materials science. We also suggest that experimental fringe data be collected with a photodiode for more elaborate data processing.

\begin{acknowledgments}
\noindent
This work was made possible by the Physics Department at the University of Ottawa. A special thanks goes out to J. Gupta for his oversight, as well as to Sylvain Robineau and to Louis Jacques for their provision of useful equipment.
\end{acknowledgments}

\section{Bibliography}

\bibliographystyle{IEEEtran}
\bibliography{citations}

\providecommand{\noopsort}[1]{}\providecommand{\singleletter}[1]{#1}%
\begin{thebibliography}{1}
\providecommand{\url}[1]{#1}
\csname url@samestyle\endcsname
\providecommand{\newblock}{\relax}
\providecommand{\bibinfo}[2]{#2}
\providecommand{\BIBentrySTDinterwordspacing}{\spaceskip=0pt\relax}
\providecommand{\BIBentryALTinterwordstretchfactor}{4}
\providecommand{\BIBentryALTinterwordspacing}{\spaceskip=\fontdimen2\font plus
\BIBentryALTinterwordstretchfactor\fontdimen3\font minus
  \fontdimen4\font\relax}
\providecommand{\BIBforeignlanguage}[2]{{%
\expandafter\ifx\csname l@#1\endcsname\relax
\typeout{** WARNING: IEEEtran.bst: No hyphenation pattern has been}%
\typeout{** loaded for the language `#1'. Using the pattern for}%
\typeout{** the default language instead.}%
\else
\language=\csname l@#1\endcsname
\fi
#2}}
\providecommand{\BIBdecl}{\relax}
\BIBdecl

\bibitem{Fizeau}
``Newton's rings,''
  \url{https://spie.org/publications/fg10_p11_newtons_rings?SSO=1}.

\bibitem{template}
G.-L.~N. et~al., ``Interference patterns produced by an evaporating droplet on
  a horizontal surface,'' \emph{Am. J. Phys.}, vol.~89, pp. 862--868, 2021.

\bibitem{background}
G.~V. . K.-P. Singh, ``Time-resolved interference unveils nanoscale surface
  dynamics in evaporating sessile droplet,'' \emph{Am. J. Phys.}, vol. 104,
  2014.

\bibitem{Laser}
``Specialty he-ne laser ml 811,''
  \url{https://i-fiberoptics.com/laser_detail.php?id=30}.

\bibitem{schlick}
\BIBentryALTinterwordspacing
C.~Schlick, ``An inexpensive brdf model for physically-based rendering,''
  \emph{Computer Graphics Forum}, vol.~13, no.~3, pp. 233--246, 1994. [Online].
  Available:
  \url{https://onlinelibrary.wiley.com/doi/abs/10.1111/1467-8659.1330233}
\BIBentrySTDinterwordspacing

\bibitem{hecht2012optics}
\BIBentryALTinterwordspacing
E.~Hecht, \emph{Optics}.\hskip 1em plus 0.5em minus 0.4em\relax Pearson, 2012.
  [Online]. Available: \url{https://books.google.ca/books?id=wcMWpBMMzIkC}
\BIBentrySTDinterwordspacing

\end{thebibliography}

\appendix
\section{Code}
Code can be provided upon request.

\section{video}

Model 1 : \href{https://www.youtube.com/watch?v=tbGZDRdzhbs}{https://www.youtube.com/watch?v=tbGZDRdzhbs}

Model 2 : \href{https://www.youtube.com/watch?v=gOzxvPE98E4}{https://www.youtube.com/watch?v=gOzxvPE98E4}

Footage :  \href{https://www.youtube.com/watch?v=HTZDUzZzZWw}{https://www.youtube.com/watch?v=HTZDUzZzZWw}

\end{document}